\documentclass[twocolumn,epjc3]{svjour3x}
\RequirePackage{fix-cm}
\RequirePackage{graphicx}
\RequirePackage{amsmath}

\newcommand{\beq}{\begin{equation}}
\newcommand{\eeq}{\end{equation}}
\newcommand{\beqa}{\begin{eqnarray}}
\newcommand{\eeqa}{\end{eqnarray}}
\def\lsim{\raise0.3ex\hbox{$<$\kern-0.75em\raise-1.1ex\hbox{$\sim$}}}
\def\gsim{\raise0.3ex\hbox{$>$\kern-0.75em\raise-1.1ex\hbox{$\sim$}}}
\def\r{{\bf r}}

\def\x{{\bf x}}
\def\y{{\bf y}}

\def\q{{\bf q}}
\def\p{{\bf p}}

\def\0{{\bf 0}}

\def\E{{\bf E}}
\def\F{{\bf F}}

\def\cal{\mathcal}

\journalname{Eur. Phys. J. C}

\begin{document}

\title{Heavy flavours in $AA$ collisions: production, transport and final
  spectra}

\author{W.M. Alberico\thanksref{infn,dip} \and A. Beraudo\thanksref{cern} \and 
A. De Pace\thanksref{infn} \and A. Molinari\thanksref{infn,dip} \and 
M. Monteno\thanksref{infn} \and M. Nardi\thanksref{infn} \and 
F. Prino\thanksref{infn} \and M. Sitta\thanksref{dipa}}

\institute{Istituto Nazionale di Fisica Nucleare, 
Sezione di Torino, via P.Giuria 1, I-10125 Torino, Italy \label{infn}
\and
Dipartimento di Fisica Teorica dell'Universit\`a di Torino, via P.Giuria 1, 
I-10125 Torino, Italy \label{dip}
\and
Physics Department, Theory Unit, CERN, CH-1211 Gen\`eve 23, Switzerland
\label{cern}
\and
Dipartimento di Scienze e Innovazione Tecnologica dell'Universit\`a del
Piemonte Orientale \\
and Gruppo Collegato INFN, Alessandria, Italy \label{dipa}
}

\date{}

\maketitle

\begin{abstract}
  A multi-step setup for heavy-flavour studies in high-energy
  nucleus-nucleus ($AA$) collisions --- addressing within a comprehensive
  framework the initial $Q\overline{Q}$ production, the propagation in the hot
  medium until decoupling and the final hadronization and decays --- is
  presented. The initial hard production of $Q\overline{Q}$ pairs is simulated
  using the POWHEG pQCD event generator, interfaced
  with the PYTHIA parton shower. Outcomes of the calculations are
  compared to experimental data in $pp$ collisions and are used as a validated
  benchmark for the study of medium effects. In the $AA$ case, the propagation
  of the heavy quarks in the medium is described in a framework provided by 
  the relativistic Lan\-gevin equation. For the latter, different choices of
  transport coefficients are explored (either provided by a perturbative
  calculation or extracted from lattice-QCD simulations) and the corresponding
  numerical results are compared to experimental data from RHIC and the LHC. In
  particular, outcomes for the nuclear modification factor $R_{AA}$ and for the
  elliptic flow $v_2$ of $D/B$ mesons, heavy-flavour electrons and non-prompt
  $J/\psi$'s are displayed.
\end{abstract}

\section{Introduction}

The purpose of our paper is to provide a comprehensive setup for the study of
heavy-flavour observables in high-energy hadronic ($pp$) and nuclear ($AA$)
collisions: $D/B$-mesons, decay electrons ($D/B\!\to\!X\,\nu\,e$) and displaced
$J/\psi$'s (i.~e., non-prompt $J/\psi$'s from $B$ decays,
$B \to J/\psi$ + X).

The interest in heavy quarks for heavy-ion phenome\-nology lies in the fact
that, being produced in the first instants and with an abundance low enough to
guarantee a small annihilation rate, they allow a tomography of the medium 
formed in high-energy $AA$ collisions, the Quark Gluon Plasma (QGP).
In fact, because of the large mass, their initial production is a
short-distance process described --- even in the $AA$ case (modulo possible 
modifications of the Parton Distribution Functions) --- by pQCD. Hence, 
differences in the final observables with respect to the $pp$ benchmark reflect 
the presence of a dense medium formed in the collision and allow us to test its 
properties. 

The last few years have seen remarkable experimental advances in heavy-flavour
measurements in $AA$ collisions, both at RHIC and at the LHC. Until very
recently experimental information on heavy quarks in high-energy nuclear
collisions was only accessible thro\-ugh the electrons from semi-leptonic
decays, without the possibility of disentangling the charm and beauty
contributions. 

Electron studies, carried out by the PHENIX~\cite{PHE} and
STAR~\cite{STARAA} experiments, were however already sufficient to
suggest a high degree of rescattering of the heavy quarks in the medium,
leading to a quenching of their decay electron spectra and to a non-vanishing
elliptic flow.

Nowadays a much richer information has become available thanks to the
measurements of $D^0,D^*$ and $D^+$ mesons in Pb-Pb collisions performed by the
ALICE experiment~\cite{ALI_D} and of displaced $J/\psi$'s detected by
CMS \cite{CMS_Jpsi}, the latter opening a new window on the understanding of
the interaction of beauty in the QCD medium. Open-charm measurements in heavy-ion
collisions were also recently presented by the STAR
collaboration~\cite{STAR_D}, based on measuring hadronic $D^0$ decays in Au-Au
collisions at RHIC. Finally, the possibility of reconstructing $D_s$ me\-sons in
a nuclear environment (preliminary results obtained by ALICE can be found
in Ref.~\cite{ALI_Ds}) will allow to study changes in the heavy-flavour
hadrochemistry arising from in-medium hadronization \cite{Kuz07,HeF13}.

On the theory side --- at variance with jet-quenching (in which one
simply considers the energy degradation of a high-$p_T$ parton) and
soft-physics studies based on hydrodynamics (in which one directly assumes to
deal with a system at local thermal equilibrium) --- heavy quark studies
require to develop tools capable of describing how off-equilibrium probes tend
asymptotically to thermalize with the surrounding environment (the Quark-Gluon
Plasma). Transport calculations provide such a tool. Various models have been
developed in the literature in the last few
years~\cite{tea,rapp,aka,aic2,BAMPS,Das11,rapp2,BAMPS2,rapp3}, differing either
in the general setup (Boltzmann, Fokker-Planck or Langevin equations) or in the
way transport coefficients are evaluated.  

The present study relies on a framework --- based on the relativistic Langevin
equation --- developed by us in the past and described in
detail in Refs.~\cite{lange0,lange}, where one can also find a comparison with
the PHENIX data available at that time. Predictions of our model for LHC energies
were provided in Refs.~\cite{qm11,hp12}. 
The calculations encompass several steps: the initial production of the heavy
$Q\overline{Q}$ pairs, taken from the next-to-leading-order pQCD event
generator POWHEG~\cite{POW} (supplemented with nuclear PDFs in the $AA$
case~\cite{eps}); the use of hydrodynamic
codes~\cite{kolb1,kolb2,rom1,rom2,rom3} to describe the evolution of the medium
from its formation up to the hadronization stage;
the propagation, governed by the Langevin equation, of the heavy quarks in
this medium; the hadronization of the heavy quarks, their decoupling from the
fireball and the simulation of the final decays. 
Transport coefficients entering into the Langevin equation and describing the
interaction of the heavy quarks with the medium have been derived within a
weak-coupling thermal field theory calculation, with proper resummations of
medium effects.

Here we update our findings, both for the $pp$ and the $AA$ cases, including
new features in the setup. Concerning the initial production, the output of the
POWHEG code for the hard event is now interfaced to the PYTHIA \cite{PYTHIA}
parton shower (the POWHEG-BOX~\cite{POWBOX} package provides the necessary
routines to accomplish this task), simulating multiple gluon radiation from the
external legs. This allows us to get, for $pp$ collisions, $p_T$-spectra of
heavy flavour hadrons and of their decay electrons in agreement with the
experimental data and also with FONLL~\cite{FONLL} calculations (often used as 
a benchmark in heavy-flavour studies), but employing a more practical tool
(since it is an event generator) providing at the same time a richer
information on the final state. This would for instance make possible to
address more differential observable, like $Q\overline{Q}$ correlations. 

For what concerns the transport coefficients, beside the perturbative values
employed in past studies, here we consider also the ones provided by recent
lattice-QCD calculations~\cite{hflat1,hflat2}. 
Though being limited to the non-relativistic/static limit, so that no
information on their momentum dependence is available, and in spite of the
well-known difficulties in extracting real-time information from euclidean
simulations, lattice results seem to indicate values of the momentum diffusion
coefficient significantly larger than perturbative predictions. Implications of
this fact on the final heavy flavour spectra will be explored in the following. 

The paper is organized as follows.
In Sect.~\ref{sec:pp} we compare POWHEG+PYTHIA results for heavy flavour
production in $pp$ collisions to experimental data obtained at RHIC and at the
LHC. 
In Sec.~\ref{sec:AA} we move to the $AA$ case: after a brief summary of the
formalism, we display results obtained with our Langevin setup using
perturbative and non-perturbative transport coefficients and we compare them to
experimental data on D mesons, heavy-flavor decay electrons and displaced
$J/\psi$'s from the STAR, PHENIX, ALICE and CMS experiments.
Finally, in Sect.~\ref{sec:concl} we draw our conclusions and suggest
perspectives for future improvements of our analysis.
    
\section{Heavy flavours in $pp$ collisions}\label{sec:pp}

Although the main goal of our paper is to study medium effects on heavy-flavour
observables in $AA$ collisions, one needs first of all to validate the tools
employed in simulating the initial $Q\overline{Q}$ production through a
comparison with the experimental data collected in $pp$ collisions. For this
purpose we rely on a standard pQCD public tool, namely POWHEG-BOX (based  
on collinear factorization), in which the hard $Q\overline{Q}$ 
event (under control, due to the large quark mass) is interfaced with a shower
stage described with PYTHIA. We start by briefly recalling the theoretical
scheme upon which it is based, since this can be of interest when addressing the
comparison of experimental data to various calculations. 

\begin{figure}
\begin{center}
\includegraphics[clip,width=0.48\textwidth]{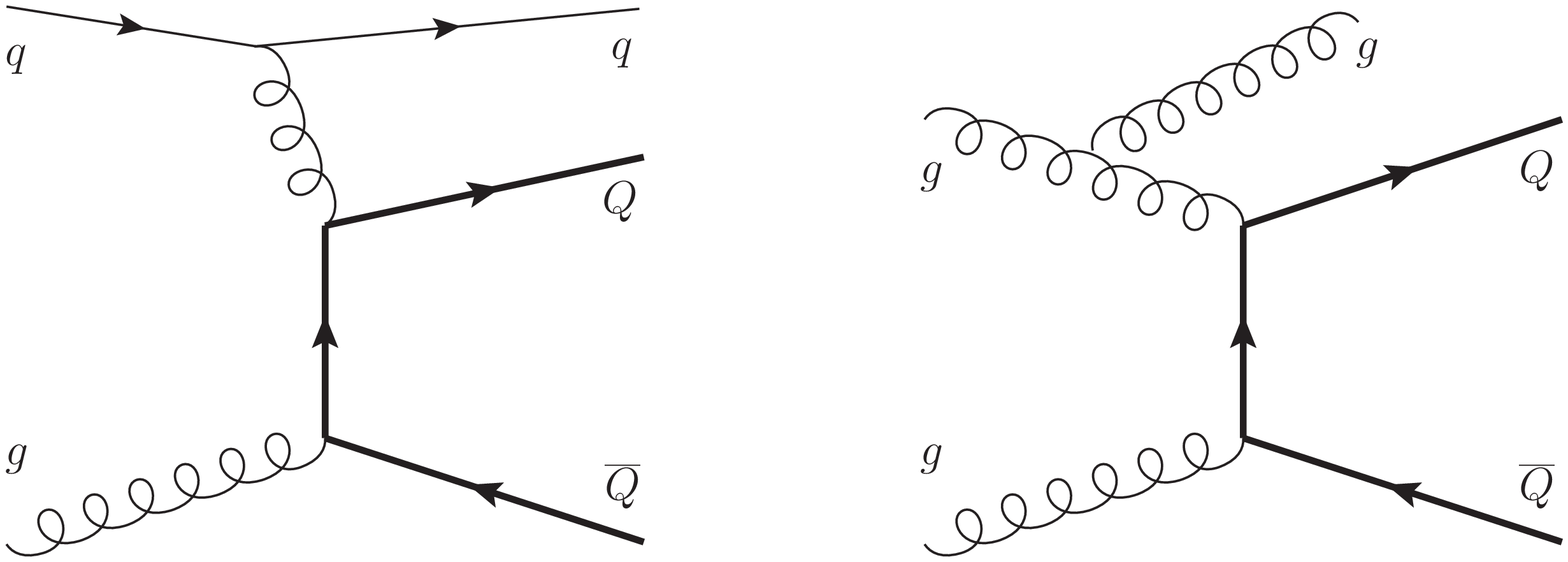}
\hskip 1cm
\includegraphics[clip,width=0.35\textwidth]{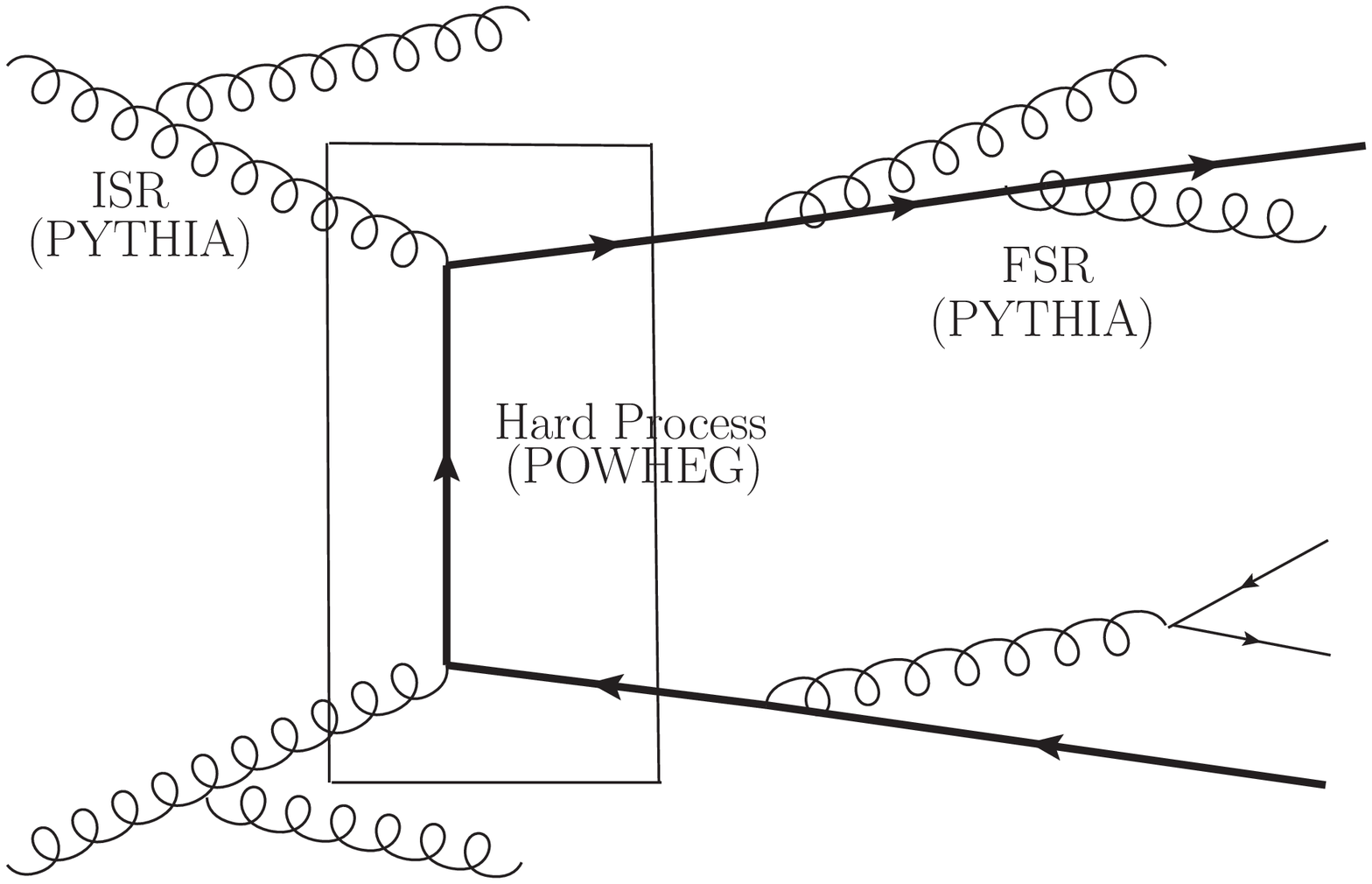}
\caption{Upper panel: some of the NLO processes contributing to the
  $Q\overline{Q}$ cross-section, e.g. $qg\to qQ\overline{Q}$ and $gg\to
  Q\overline{Q}g$. Lower panel: the structure of a typical event arising from
  the POWHEG-BOX setup, with the hard process (in the box) followed by a shower
  stage.}\label{fig:NLO} 
\end{center}
\end{figure}

The large mass of $c$ and $b$ quarks makes their partonic production
cross-section accessible to pQCD calculations. 
Tools like POWHEG \cite{POW} and MC@NLO \cite{MC@NLO} generate hard
$Q\overline{Q}$ events according to the pQCD cross-section at
next-to-leading-order (NLO) accuracy and represent the state of the art as 
event generators. Examples of NLO processes contributing to the $Q\overline{Q}$
hadro-production cross-section are shown in the upper Fig.~\ref{fig:NLO}. The
hard pairs are then showered with PYTHIA 6.4 \cite{PYTHIA}, benefiting from
the user-friendly interface provided by the POWHEG-BOX package~\cite{POWBOX}:
this allows us to include the effects of Initial State Radiation (ISR) and
Final State Radiation (FSR), resumming multiple emission of soft/collinear
gluons at Leading Log (LL) accuracy. Intrinsic-$k_T$ corrections (with $\langle
k_T^2\rangle\!=\!1$ GeV$^2$) are also included in the simulation of the
heavy-quark production.
The structure of a typical event resulting from the above chain is displayed in
the lower panel of Fig.~\ref{fig:NLO}. 

\begin{figure}
\begin{center}
\includegraphics[clip,height=2cm]{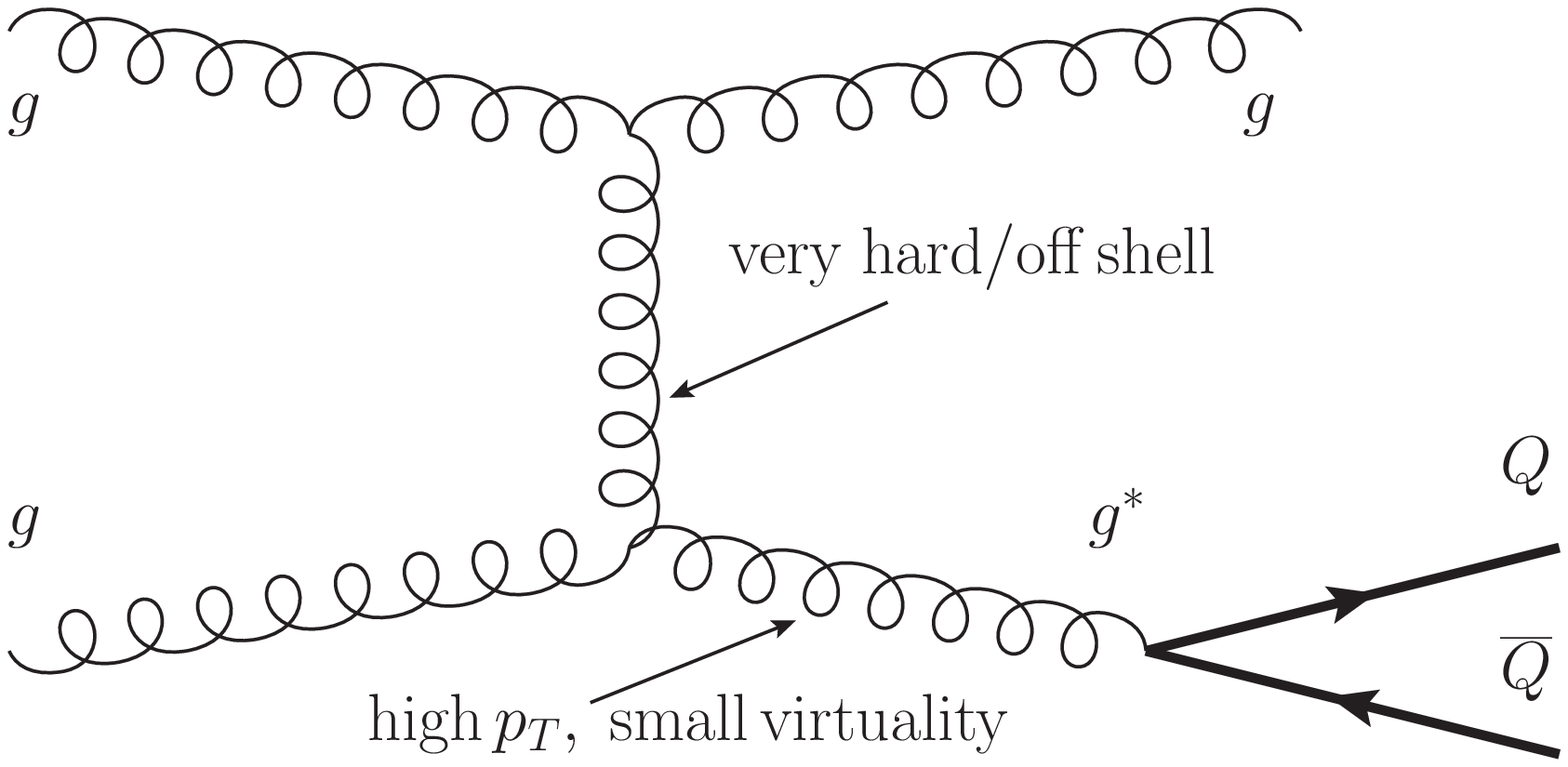}
\includegraphics[clip,height=2cm]{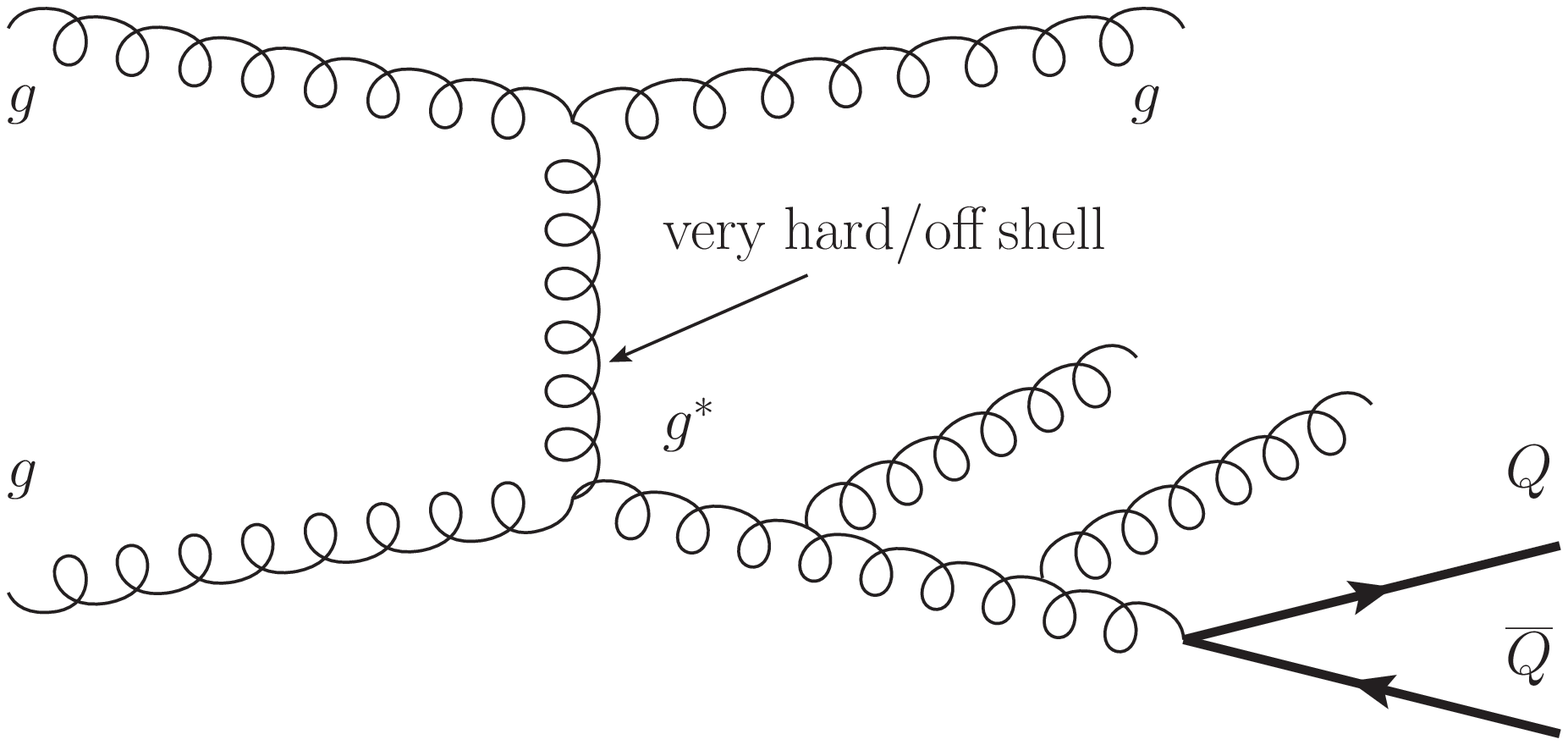}
\caption{Upper panel: a NLO contribution to $Q\overline{Q}$ production from
  gluon-splitting. Lower panel: an example of process accounted for by the
  FONLL calculation with a $g\!\to\!Q\overline{Q}$ splitting at the end of the
  high-$p_T$ gluon DGLAP evolution.}\label{fig:splitting} 
\end{center}
\end{figure}

We now wish to illustrate how the above setup compares with the FONLL (Fixed
Order Next to Leading Log) calculation of the inclusive charm/beauty production
cross-section~\cite{FONLL}, often regarded as the standard theoretical
benchmark in experimental heavy-flavour analysis. 
A particular NLO contribution comes from 
processes as the one displayed in Fig.~\ref{fig:splitting}, in which an
intermediate gluon with virtuality much smaller than the (high) $p_T$ of the
hard scattering splits into a $Q\overline{Q}$ pair. The process can be seen as
the convolution of the cross-section for the production of an almost on-shell
gluon ($gg\!\to\! gg^*$) followed by its splitting into a $Q\overline{Q}$ pair
($g^*\!\to\! Q\overline{Q}$). 
Integrating over the possible virtuality of the gluon from the
$Q\overline{Q}$-threshold up to $p_T$ one gets that (at NLO) the average
$Q\overline{Q}$ multiplicity inside a gluon-jet is of the order of
\begin{displaymath} 
N(Q\overline{Q})\sim\frac{\alpha_s}{6\pi}\ln\frac{p_T^2}{m_Q^2},
\end{displaymath}
so that for very large $p_T$ one can have $\alpha_s\ln(p_T/M)\!\sim\!1$ and a
Fixed Order Calculation becomes no longer meaningful. One has then to resum the 
full DGLAP evolution of the gluon from the hard interaction up to its splitting
into a $Q\overline{Q}$ pair, as shown in the lower panel of
Fig.~\ref{fig:splitting}. This goes beyond the processes accounted for by NLO
pQCD event generators, but is included in the FONLL calculation, which ---
employing NLO Altarelli-Parisi splitting functions --- allows one to reach Next
to Leading Log accuracy, resumming all $\alpha_s^n\ln^n(p_T/M)$ and
$\alpha_s^{n+1}\ln^n(p_T/M)$ terms.

We shall see below that the differences between the resummation schemes adopted
by POWHEG and FONLL give rise to differences between the corresponding heavy
quark spectra well within the theoretical uncertainties and, generally, in
fairly good agreement with the available experimental data.

Concerning the hadronization stage we adopt essentially the same fragmentation
setup employed by FONLL, which was tuned by the authors to reproduce
experimental $e^+e^-$ data.  
Heavy quarks are made hadro\-nize by sampling different hadron species from $c$
and $b$ fragmentation fractions extracted from experimental
data \cite{ZEUS,ALEPH1,HFAG}. Then, hadron momenta are sampled from
Fragmentation Functions (FFs) calculated in heavy-quark effective theory
(HQET)~\cite{Braaten}, which entails a dependence on the parameter $r$. 
For charm quarks, we have used, for the $r$ parameter, the
value fitted (in particular, fixing the higher moments of the FF) in the
FONLL framework~\cite{FONLL_D_Tevatron} to ALEPH data~\cite{ALEPH1} at the 
LEP $e^{+}e^{-}$ collider (i.e., $r=0.06$ for $m_c=1.3$~GeV).
For bottom fragmentation we used the functional form proposed by
Kartvelishvili et al.~\cite{KAR}, whose single parameter $\alpha$ was fitted in 
the FONLL framework \cite{FONLL_B_Tevatron} to ALEPH~\cite{ALEPH2} and
SLD~\cite{SLD} $e^{+}e^{-}$ data (namely $\alpha\!=\!29.1$ for
$m_{b}\!=\!4.75$~GeV).

In order to estimate the systematic uncertainties associated to different FF
choices, we have re-done the calculation with an alternative choice for the
fragmentation function, namely using for the parameter $r$ the definition 
proposed in Ref.~\cite{Braaten}, which was $r\equiv(m_{H}-m_{Q})/m_{H}$  ($m_H$
and $m_Q$ being the hadron and the heavy quark masses, respectively), 
resulting, e.~g., in $r=0.3$ for $D^0$ and $D^+$ mesons (using $m_c=1.3$~GeV).
\begin{figure}
\begin{center}
\includegraphics[clip,width=0.48\textwidth]{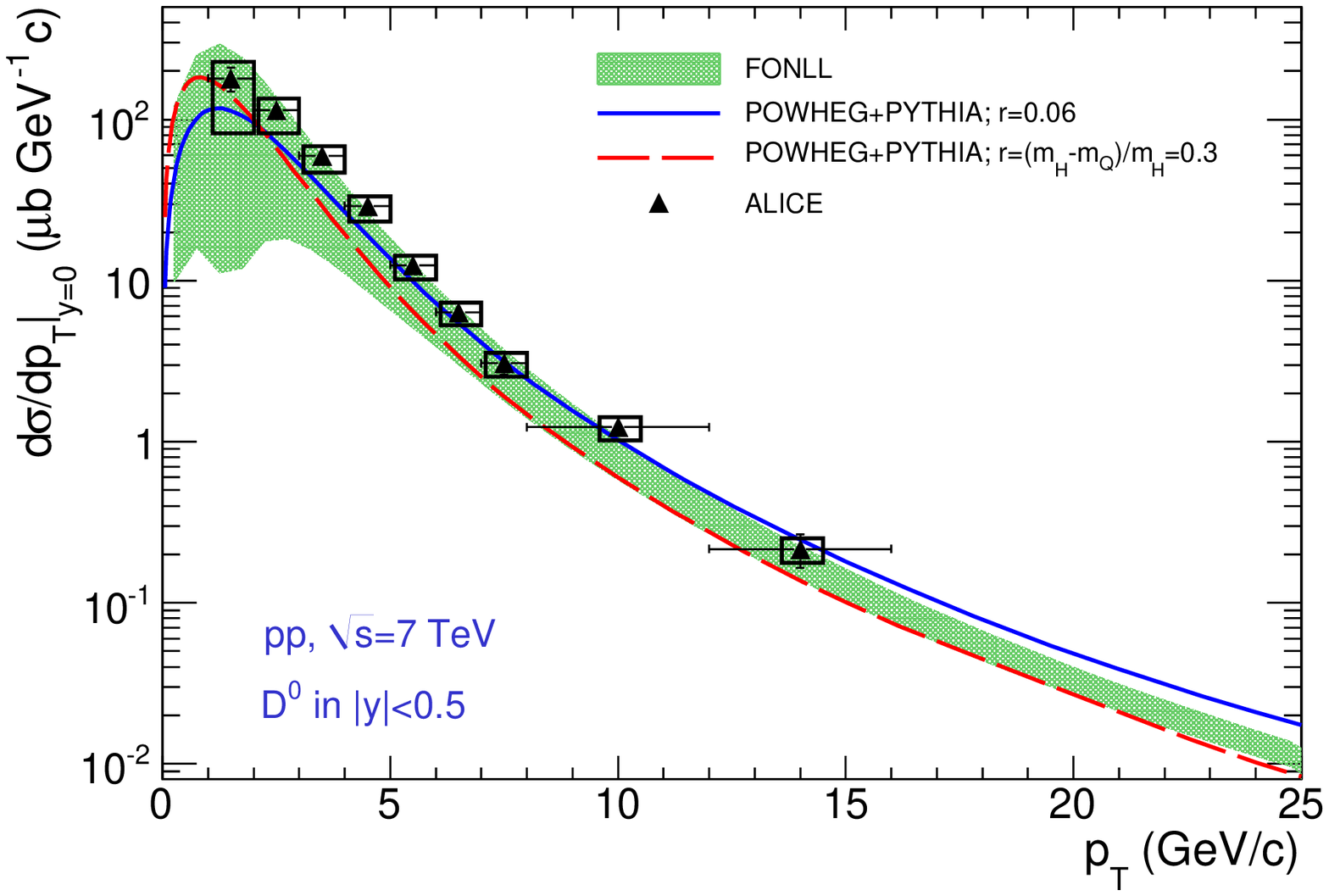}
\caption{POWHEG+PYTHIA predictions for $D^0$ meson spectra at 
  $\sqrt{s}\!=\!7$~TeV (with different parameter choices for the fragmentation
  stage) compared to ALICE data \cite{ALIpp7} and to the FONLL systematic 
  uncertainty band.}
\label{fig:D7} 
%\end{center}
%\end{figure}
%\begin{figure}[b]
%\begin{center}
\includegraphics[clip,width=0.48\textwidth]{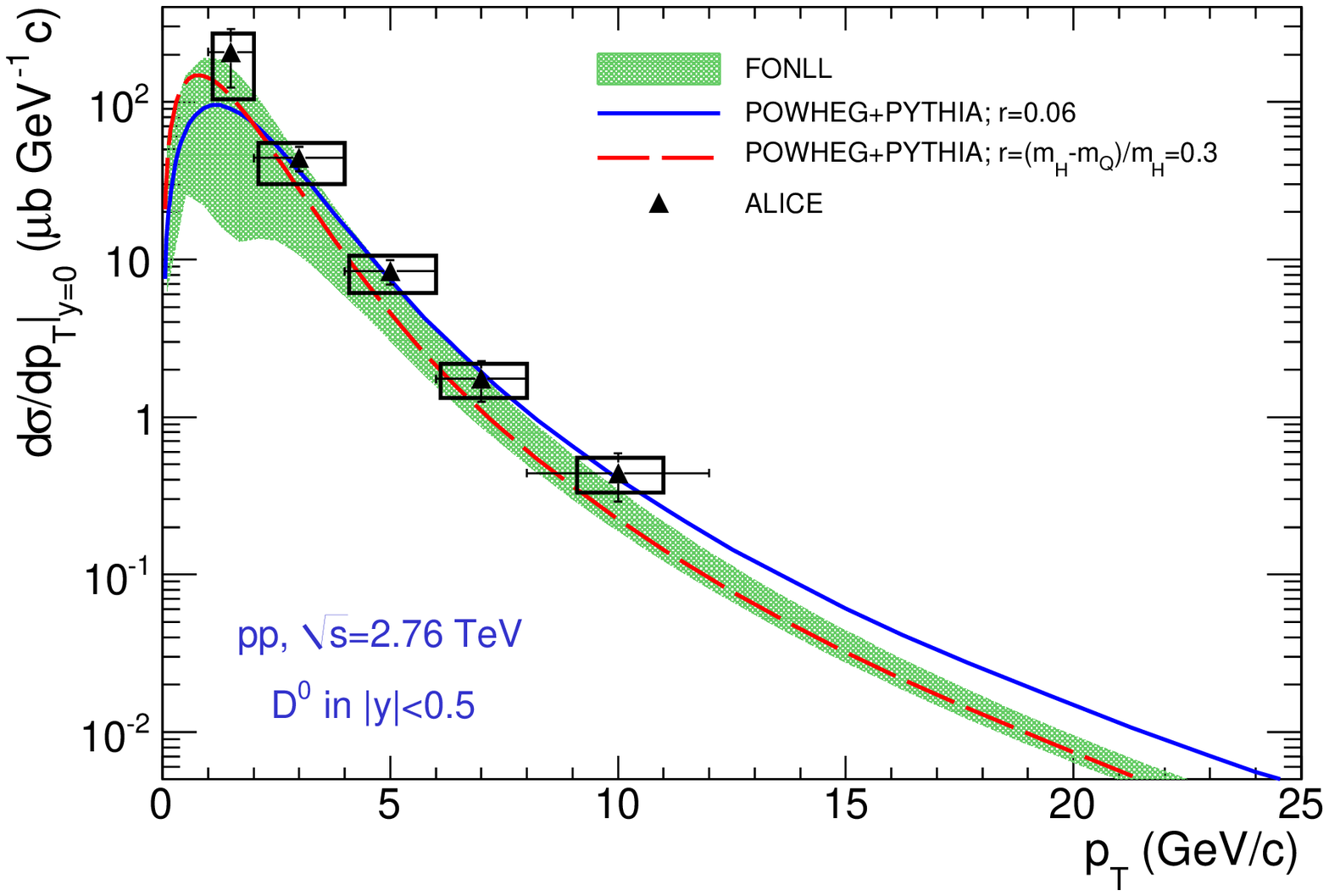}
\caption{POWHEG+PYTHIA predictions with for $D^0$ spectra at
  $\sqrt{s}\!=\!2.76$ TeV (with different fragmentation schemes) compared with
  ALICE data \cite{ALIpp2}.}
\label{fig:D2} 
\end{center}
\end{figure}

In Fig.~\ref{fig:D7} we display the outcomes of the
POWHEG+ PYTHIA setup for $D^0$ mesons in $pp$ collisions compared to 
ALICE \cite{ALIpp7} data at $\sqrt{s}=7$~TeV, together with the FONLL
uncertainty band \cite{Cacciari:2012ny}. 
Both data and theoretical predictions include the ``primary'' production ($c\to
D$) as well as the $D^*$ feed-down ($c\to D^*\to D$). 
In the POWHEG+PYTHIA setup we have kept the default values for the
renormalization and factorization scales and for the charm mass we have set
$m_c=1.3$~GeV.
The results displayed have been obtained with two different values of
the $r$ parameter ($r=0.3$, entailed by the HQET calculation, and $r=0.06$,
fitted to $e^+e^-$ data). 
As it can be seen, both the POWHEG+PYTHIA outcomes lie very close to the FONLL
uncertainty band. Experimental data lie at the upper edge of the FONLL
theoretical  uncertainty band and are nicely described by choosing $r=0.06$. 
Analogous results hold for the other open-charm mesons ($D^+$ and $D^*$)
reconstructed by ALICE.

We have also tested POWHEG+PYTHIA outcomes at $\sqrt{s}\!=\!2.76$ TeV, which
will represent the $pp$ benchmark for the study of medium effects in $AA$
collisions at the same center-of-mass energy. In Fig.~\ref{fig:D2}, referring
to the case of $D^0$ mesons, one can compare theory points to ALICE
experimental data~\cite{ALIpp2}. 
Also here the setup with $r=0.06$ gives a good description of the data. Hence,
in the rest of the paper our default choice for charm production and
hadronization will be $m_c=1.3$~GeV and the HQET FF with $r=0.06$.

\begin{figure}
\begin{center}
\includegraphics[clip,width=0.48\textwidth]{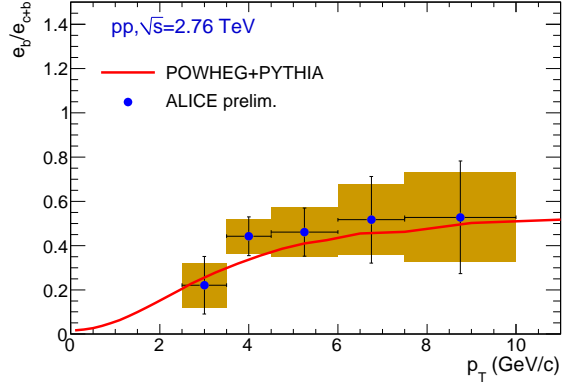}
\caption{The ratio of electrons $e_b$ from beauty decays over the total number
  of non-photonic electrons $e_{c+b}$ in $pp$ collisions at $\sqrt{s}\!=\!2.76$
  TeV at the LHC. POWHEG+PYTHIA results are compared to preliminary ALICE
  data~\cite{ALIhfepp}. }
\label{fig:eb/ec+eb}  
\end{center}
\end{figure}

\begin{figure}[t]
\begin{center}
\includegraphics[clip,width=0.48\textwidth]{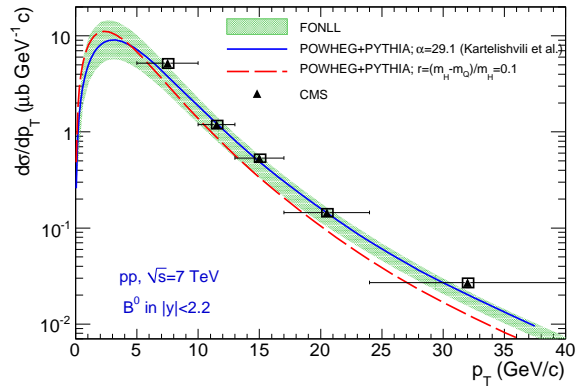}
\caption{POWHEG+PYTHIA predictions for $B^0$ meson spectra at 
  $\sqrt{s}\!=\!7$~TeV (with different fragmentation schemes) compared to
  CMS data~\cite{CMSppB} and to the FONLL systematic uncertainty band.}
\label{fig:B7} 
\end{center}
\end{figure}

For completeness, in Fig.~\ref{fig:eb/ec+eb} we also plot, for the same 
center-of-mass energy, the ratio of electron spectra $e_b$ from semileptonic 
decays of beauty over the inclusive $e_{c+b}$ one. The POWHEG+PYTHIA setup 
predicts that the beauty contribution increases, becoming as large as the charm
contribution at $p_T\!\sim\!7$ GeV, in agreement with the ALICE preliminary
data~\cite{ALIhfepp}.

\begin{figure}
\begin{center}
\includegraphics[clip,width=0.48\textwidth]{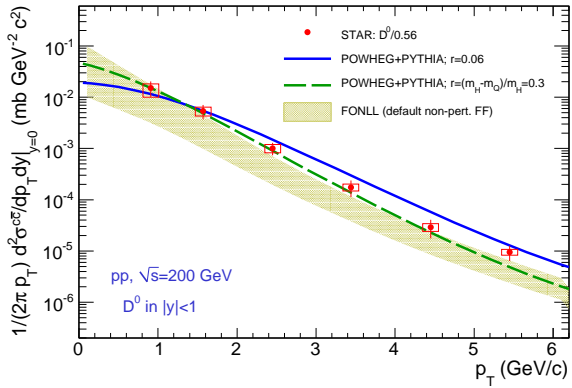}
\caption{$D^0$ $p_T$-differential cross-section in $pp$ collisions at
  $\sqrt{s}\!=\!200$ GeV, measured by the STAR collaboration~\cite{STARppD}, 
  compared to the predictions of the POWHEG+PYTHIA setup employed in the
  present work; note that the curves are normalized to the $c\bar{c}$ production
  cross-section.}
\label{fig:DppSTAR}
\end{center}
\end{figure}

In Fig.~\ref{fig:B7} we address beauty production, comparing results for $B^0$
mesons to CMS data~\cite{CMSppB}.
Again, in the POWHEG+PYTHIA setup we have kept the default values for the
renormalization and factorization scales and we have used for the bottom mass 
$m_b=4.8$~GeV. Two different fragmentation functions are tested: the one 
provided by HQET and the parametrization proposed by Kartvelishvili et al., 
fitted to $e^+e^-$ data.
The latter is found to describe pretty well the data and will be the one
employed in the rest of the paper. An analogous degree of agreement has been 
found for $B^+$ and $B_s^0$ mesons.

Finally, in Fig.~\ref{fig:DppSTAR} we display the outcomes of the
POWHEG+PYTHIA setup (with the same parameters used for the LHC case) for $D^0$
mesons in $pp$ collisions at $\sqrt{s}\!=\!200$~GeV compared to
STAR~\cite{STARppD} data and to the FONLL uncertainty band.
As it can be seen, our default choice is now slightly overshooting the data at
$p_T$'s in the 2--6~GeV/c range, which appear to be better described by r$=0.3$.
Note, however, that we have not tried any adjustment of the scales in the POWHEG
setup. 

To summarize, we have shown how --- within the $p_T$ range of interest for
heavy-flavour studies at the LHC --- the POWHEG+PYTHIA setup, which will be
employed in the rest of the paper, provides results in 
quantitative agreement with the available experimental data and also with
FONLL; with respect to the latter, it is for our purposes a more practical tool,
providing full information on the event, of potential interest for future less
inclusive measurements, such as $Q\overline{Q}$ correlations. 
     
\section{Heavy flavours in $AA$ collisions}\label{sec:AA}

In this section we discuss the results of our transport setup for
several heavy-flavour observables accessible in $AA$ collisions at RHIC and at
the LHC. All the experimental data so far available (namely, non-photonic
electrons measured by PHENIX and ALICE, $D$ mesons reconstructed by ALICE
and STAR, heavy flavour decay muons measured by ALICE at forward rapidity, and
displaced $J/\psi$'s from $B$ decays measured by CMS) signal a high degree of
rescattering of the heavy quarks in the medium formed in heavy-ion collisions. 
The challenge for the theoretical calculations is then to reproduce within a
coherent setup the rich amount of heavy-flavour data nowadays available, in
particular the quenching of the $p_T$ spectra --- commonly studied via the
nuclear modification factor ($R_{\rm AA}$) --- and the elliptic flow $v_2$.

At very large $p_T$ --- the mass playing a negligible role --- there is no
reason to believe that the energy-loss mechanism of $c$ and $b$ quarks should
be different from the one of light quarks, usually described by medium-induced
gluon radiation; their final semileptonic decays are even proposed (and already
used) as a tool to tag quark jets in a heavy-ion environment, shedding light on
jet-quenching in $AA$ collisions. 

On the other hand, at small or moderate values of $p_T$ the large quark mass is
expected to suppress the rate of gluon radiation, favoring the transition to a
regime where collisional energy loss (in particular for beauty) plays the major
role. Furthermore, in such a $p_T$ range it becomes mandatory to employ tools
(\emph{transport calculations}) capable of describing the asymptotic relaxation
of heavy quarks to thermal equilibrium. 
Actually, in the case of a static medium heavy quarks --- sooner or later ---
would always thermalize, no matter how strongly they are coupled with the
plasma. The fact that in the actual experimental situation
--- of a medium with a finite life-time and a non-vanishing expansion rate ---
heavy quarks are found to follow the collective flow of the fireball can put
tight constraints on their relaxation time to thermal equilibrium.

In our approach the dynamics of heavy quarks in the QGP is studied through the
relativistic Langevin equation. The general setup has been presented in
previous publications~\cite{lange0,lange} and its predictions were already
compared with RHIC and first LHC data~\cite{qm11,hp12}. Here we extend the
setup along different directions. While in previous studies we relied on a
microscopic derivation of the heavy-flavour transport coefficients performed
within a weakly-coupled scenario, here we also test the predictions obtained
with the values provided by recent non-perturbative lattice-QCD
simulations. 

Furthermore we will put a stronger emphasis on beauty measurements
(already feasible via displaced $J/\psi$ detection by CMS and at the center of
the ALICE upgrade program) and on their potentially major role in getting
information on the quark-medium coupling. The large mass of $b$ quarks makes in
fact a description of their energy-loss (and thermal relaxation) in the plasma
in terms of uncorrelated random collisions working over an extended
$p_T$-range; moreover, non-perturbative information on heavy-quark transport
coefficients arising from lattice-QCD simulations performed in the static
($M\!\to\!\infty$) limit, if questionable for charm, may provide a good
guidance in the case of beauty. 

Finally, at variance with the case of charm, for $b$ quarks hadronization
should play a minor role as a source of systematic uncertainty. While in
elementary collisions it occurs via 
\emph{fragmentation}, with the final hadron carrying away a fraction $z$ of the
parent parton momentum, in $AA$ collisions it has been pointed out that a
\emph{coalescence} mechanism (in which a hard parton hadronizes picking-up a
companion quark/antiquark from the thermal bath) might explain several features
of ha\-dron production at moderate $p_T$. In particular, the latter mechanism
entails a momentum gain in the hadroni\-zation stage. 

Medium modification of hadronization, while being an interesting research topic 
in itself, represents an important source of uncertainty in our studies, 
which are aimed at getting information on what occurred in the partonic phase.
Coalescence can in fact modify hadron spectra both at the level of their shape
(entailing a momentum gain, at variance with fragmentation) and of their absolute
normalization, due to possible changes in the hadrochemistry (as suggested for
instance by the recent ALICE $D_s$ measurements~\cite{ALI_Ds}). 
This might have a minor relevance in the case of beauty. 
Since its vacuum fragmentation function is very hard and the momentum-gain in 
case of coalescence is very small (of order $(m_q/M_b)p_T^b$), the two opposite
scenarios should not entail dramatic differences for the final hadron spectra. 
Furthermore, measurements of displaced $J/\psi$'s should be less sensitive to
changes in the relative abundances of the parent beauty hadrons. In summary:
beauty measurements, being less affected by uncertainties due to the
hadronization mechanism, can potentially provide a cleaner information on the
properties of the medium and of the coupling of external probes to the latter. 

\begin{figure*}[!t]
\begin{center}
\includegraphics[clip,width=0.9\textwidth]{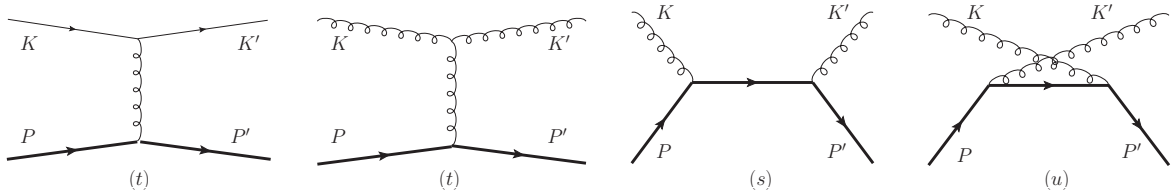}
\caption{Diagrams for the hard scattering of a heavy quark off a light
  (anti-)quark and a gluon from the medium.}   
\label{fig:quarkgluon}
\end{center}
\end{figure*}

\subsection{The relativistic Langevin equation}

Within the Langevin framework, the time evolution of the momentum of a
relativistic Brownian particle is provided by the following stochastic
differential equation 
\begin{equation}
  \frac{\Delta{\bf p}}{\Delta t}=-\eta_D(p){\bf p}+\vec{\xi}(t),
\label{eq:lange_r_d}
\end{equation}
where the \emph{drag coefficient} $\eta_D(p)$ describes the deterministic
friction force acting on the heavy quark, whereas the term $\vec{\xi}$ accounts
for the random collisions with the constituents of the medium. 
The effect of the stochastic term is completely determined once its temporal
correlation function is fixed. The latter is usually assumed to be given by 
\begin{equation}
  \langle\xi^i(t)\xi^j(t')\rangle=b^{ij}({\bf p})\delta_{tt'}/\Delta t,
\label{eq:noise1}
\end{equation}
entailing that collisions at different time-steps are uncorrelated. The tensor
$b^{ij}({\bf p})$ can be decomposed with a standard procedure according to
\begin{equation}
  b^{ij}({\bf p})\equiv \kappa_L(p)\hat{p}^i\hat{p}^j+\kappa_T(p)
(\delta^{ij}-\hat{p}^i\hat{p}^j),
\label{eq:noise2}
\end{equation}
with the coefficients $\kappa_{L/T}(p)$ representing the squared
longitudinal/transverse momentum per unit time exchanged by the quark with the 
medium. Finally, the drag coefficient $\eta_D(p)$ is fixed in order to fulfill
equilibrium: for large times the momenta of an ensemble of 
heavy quarks should approach a thermal Maxwell-J\"uttner
distribution. This request leads to the relativistic generalization of the
Einstein relation 
\beq
 \eta_D(p)\equiv\frac{\kappa_L(p)}{2TE_p}+{\rm discr.\;corr.}, 
\eeq
where the corrections on the right hand side are fixed in order to ensure that
in the continuum, $\Delta t\!\to\! 0$, limit Eq.~(\ref{eq:lange_r_d}) reduces to
the same Fokker-Planck equation, independently on the discretization scheme
employed. At each time step the update of the quark momentum (and position) has
to be performed in the local fluid rest frame (for more detail on the procedure
see Ref.~\cite{lange}), where the transport coefficients $\kappa_{L/T}(p)$ and
$\eta_D(p)$ are defined. They can be in principle obtained from first-principle
calculations and represent the key quantities to establish a link between
the underlying microscopic theory (QCD) and the final observables accessible in
$AA$ collisions. Their evaluation will be the subject of the next section. 

\subsection{The transport coefficients: weak-coupling vs non-perturbative
  results} 

Heavy quark transport coefficients can be evaluated starting from their
definition 
\beq
\kappa_L=\left\langle\frac{\Delta \q_L^2}{\Delta t}\right\rangle\qquad{\rm and}\qquad \kappa_T=\frac{1}{2}\left\langle\frac{\Delta \q_T^2}{\Delta t}\right\rangle.
\eeq
We will consider the outcomes of two different approaches: weak-coupling
calculations and lattice-QCD simulations.  

\begin{figure}[b]
\begin{center}
\includegraphics[clip,width=0.48\textwidth]{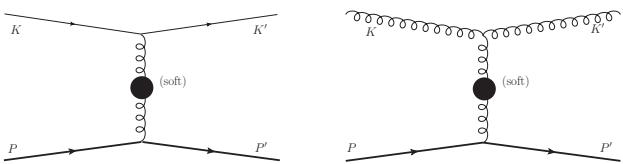}
\caption{Scattering mediated by long-wavelength gluons require the resummation
  of medium effects.}
\label{fig:soft_scatt}
\end{center}
\end{figure}

The momentum broadening (and degradation) of heavy quarks in the medium must
arise from their interaction with the other constituents of the plasma: light
quarks and gluons. Within a perturbative setup, the lowest order diagrams to
be considered are the ones in Fig.~\ref{fig:quarkgluon}. If the four-momentum
exchange is sufficiently hard ($|t|\!>\!|t|^*$, where
$t\!\equiv\!\omega^2-\q^2$) one is dealing with a short-distance process and
the result is given by a kinetic pQCD calculation: 
\begin{multline}
\kappa_{L,{\rm hard}}^{g/q} =
\frac{1}{2E}\!\!\int_k\frac{n_{B/F}(k)}{2k}\!\!\int_{k'} 
\frac{1\pm n_{B/F}(k')}{2k'}\!\!\int_{p'}\frac{\theta(|t|-|t|^*)}{2E'}\\ 
\times(2\pi)^4\delta^{(4)}(P+K-P'-K')\left|\overline{{\cal
    M}}_{g/q}(s,t)\right|^2 \q_L^2
\label{eq:klhard}
\end{multline}
and
\begin{multline}
\kappa_{T,{\rm hard}}^{g/q} =
\frac{1}{2E}\!\!\int_k\frac{n_{B/F}(k)}{2k}\!\!\int_{k'}\frac{1\pm
  n_{B/F}(k')}{2k'}\!\!\int_{p'}\frac{\theta(|t|-|t|^*)}{2E'}\\  
\times(2\pi)^4\delta^{(4)}(P+K-P'-K')\left|\overline{{\cal
    M}}_{g/q}(s,t)\right|^2 \frac{\q_T^2}{2}.
\label{eq:kthard} 
\end{multline}
If on the contrary the momentum transfer is soft ($|t|\!<\!|t|^*$), the
scattering involves the exchange of a long wavelength gluon, which requires the
resummation of medium effects, as displayed in Fig.~\ref{fig:soft_scatt}. This
can be done in hot-QCD within the Hard Thermal Loop approximation. The
corresponding contribution to $\kappa_{L/T}$ has been derived and evaluated in
Refs.~\cite{lange0,lange}, where the interested reader can find more
details. Eventually, one has to sum-up the soft and hard contributions to the
transport coefficients 
\beq
\kappa_{L/T}=\kappa_{L/T}^{\rm soft}+\kappa_{L/T}^{\rm hard},
\eeq 
\begin{figure}
\begin{center}
\includegraphics[clip,width=0.48\textwidth]{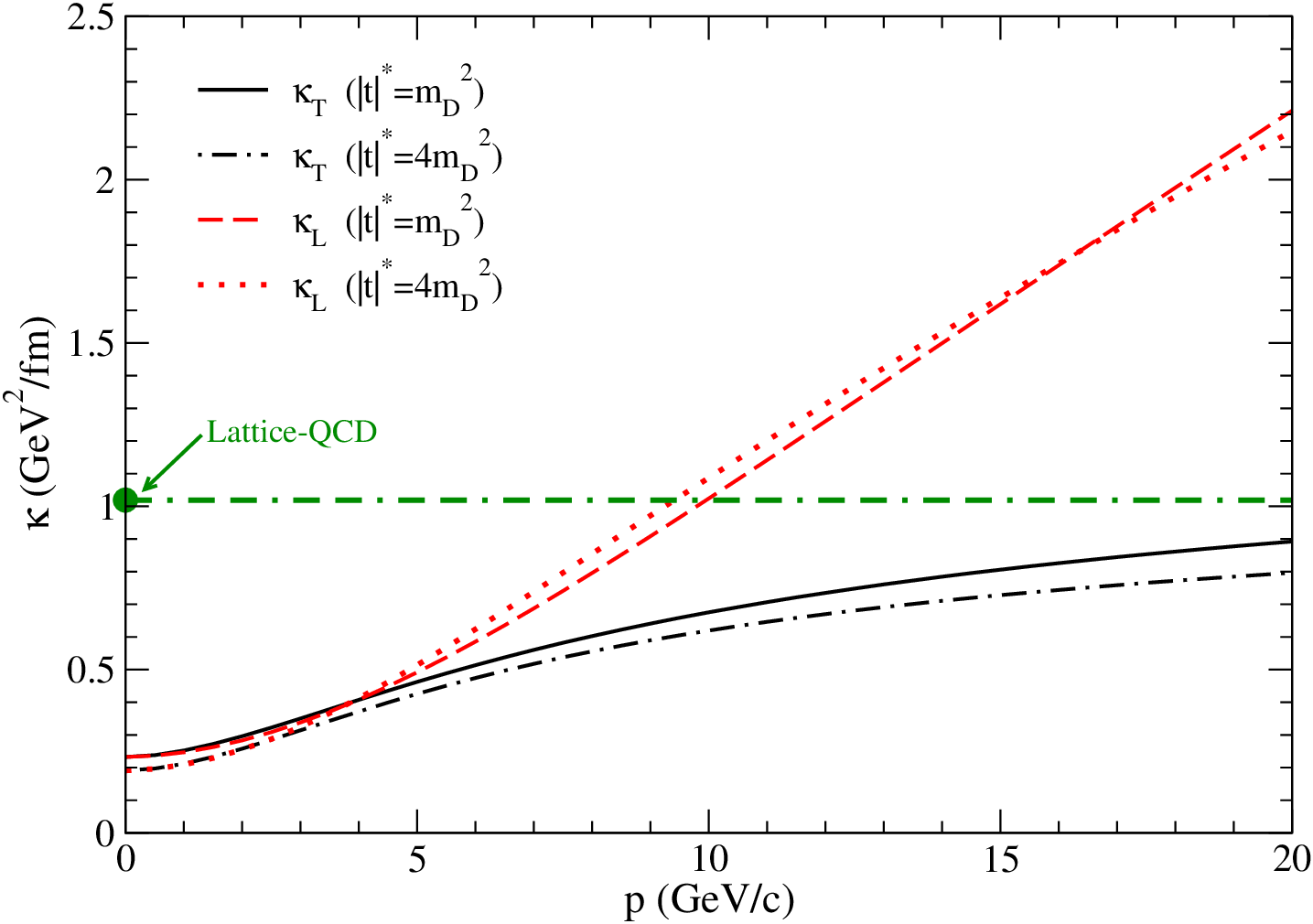}
\caption{Transport coefficients for $b$ quarks in the QGP. Weak coupling
  (HTL+pQCD) results for $\kappa_{L/T}(p)$ are compared to the data provided 
  by the lattice-QCD simulations of Ref.~\cite{hflat1}.
  The curves refer to the temperature $T=400$~MeV and, for the HTL+pQCD 
  coefficients, the running coupling has been evaluated at the scale 
  $\mu\!=\!1.5\pi T$ in the soft contribution and at the scale
  $\mu\!=\!\sqrt{|t|}$, set by the four-momentum exchange, in the calculation
  of the hard collisions. Note that the lattice-QCD prediction is just given by
  the dot at $p=0$, the straigth line representing the value used at finite $p$
  for lack of momentum dependence.}
\label{fig:kappa_b_lat}  
%\end{center}
%\end{figure}
%\begin{figure}
%\begin{center}
\includegraphics[clip,width=0.48\textwidth]{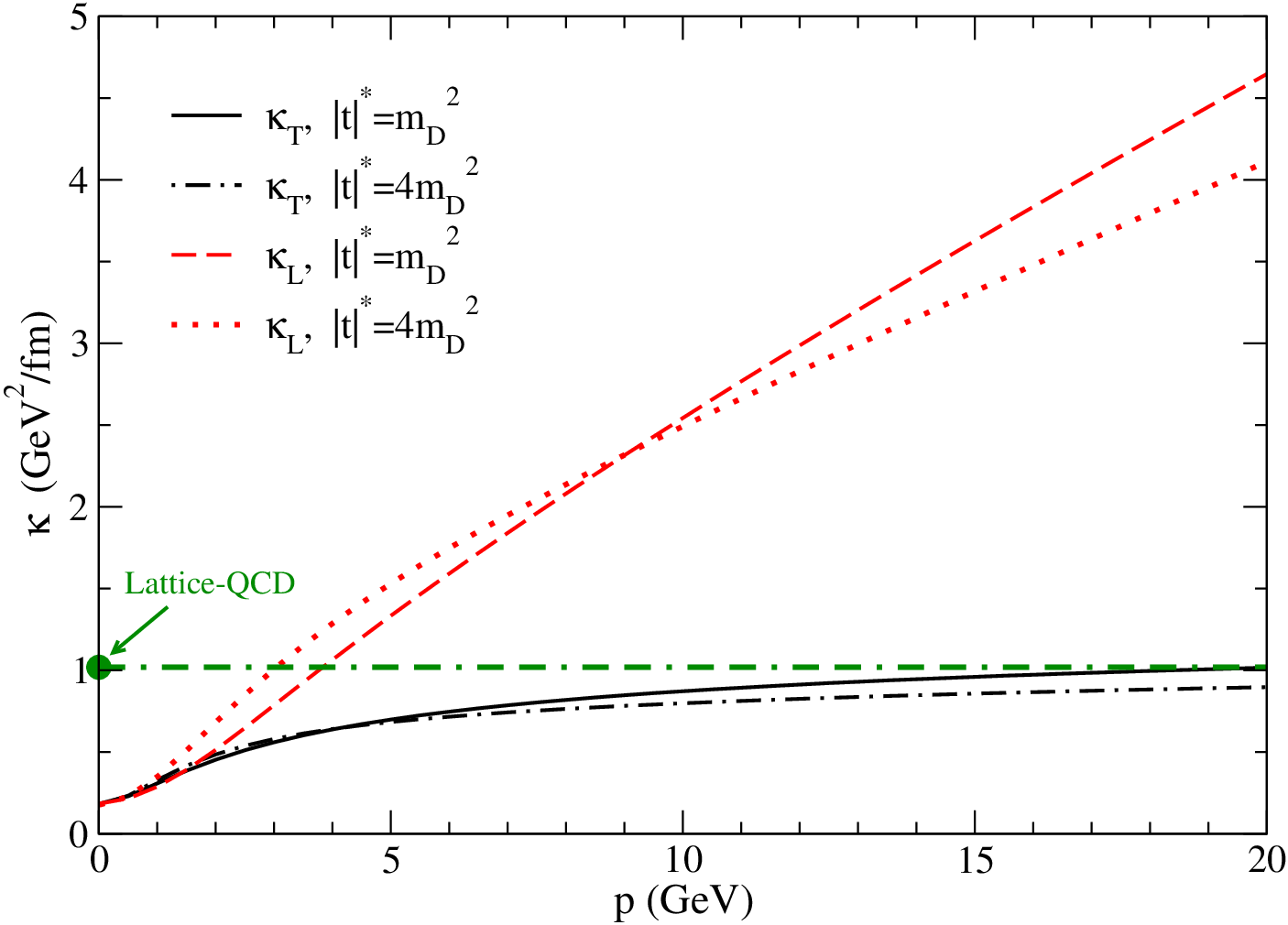}
\caption{The same as in Fig.~\ref{fig:kappa_b_lat}, but for $c$ quarks.}
\label{fig:kappa_c}
\end{center}
\end{figure}
checking that the final result is not too sensitive to the artificial
intermediate cutoff $|t|^*$: choosing $|t|^*\!\sim\!m_D^2$ (the Debye-mass
$m_D$ being responsible for the screening of electric fields in the plasma) one
verifies that this is actually the case~\cite{lange}, as it can be seen in
Figs.~\ref{fig:kappa_b_lat} and \ref{fig:kappa_c}.
Concerning the scale of the strong coupling constant $g$, the latter has been
set at the typical thermal momentum, $\mu\!\sim\!T$, in the soft contribution 
and to the squared-momentum transfer in the collisions,
$\mu\!\sim\!\sqrt{|t|}$, in the evaluation of $\kappa^{\rm hard}$.  

An independent way to extract the transport coefficients from the underlying
microscopic theory comes from lattice-QCD simulations. The results we shall
employ in the calculations have been obtained in the static ($m_Q\!\to\!\infty$)
limit and refer to the momentum diffusion coefficient $\kappa$. The latter is
calculated~\cite{sola,lai} starting from the non-relativistic limit of the
Langevin equation (here written in the continuum limit):  
\beq
\frac{dp^i}{dt}=-\eta_D p^i+ \xi^i(t),\quad{\rm with}\quad
\langle\xi^i(t)\xi^j(t')\rangle\!=\!\delta^{ij}\delta(t-t')\kappa. 
\eeq 
Hence, in the $p\!\to\!0$ limit, $\kappa$ reduces to the evaluation of the
following force-force correlator: 
\beqa
\kappa&=&\frac{1}{3}\int_{-\infty}^{+\infty}dt\langle\xi^i(t)\xi^i(0)\rangle_{\rm
  HQ} \nonumber \\ 
&\approx&\frac{1}{3}\int_{-\infty}^{+\infty}dt\langle F^i(t)F^i(0)\rangle_{\rm
  HQ},\label{eq:force-force} 
\eeqa
where the expectation value is taken over a thermal ensemble of states
containing one Heavy Quark (HQ) (further details are provided
in~\ref{app:kappa}). 
In the static limit (magnetic effects being negligible) the force is nothing
but the colour-electric field acting on the heavy quark, namely: 
\beq
\F(t)=g\int\! d\x\, Q^\dagger(t,\x)t^aQ(t,\x)\E^a(t,\x),\label{eq:electric}
\eeq
where $Q$ and $Q^\dagger$ are non-relativistic fields destroying and creating a
heavy quark respectively.
In Refs.~\cite{hflat1,hflat2} $\kappa$ has then been extracted from euclidean
electric-field correlators, getting, within the explored temperature range,
$\kappa/T^3\sim 2.5\div4$. For our calculations we shall rely on a linear 
interpolation of the values of $\kappa(T)$ quoted in Ref.~\cite{hflat1}. 
The friction and spatial-diffusion coefficients are then set through the 
Einstein relation to $\eta_D\!=\!\kappa/2E_pT$ and $D\!=\!2T^2/\kappa$. 

Figs.~\ref{fig:kappa_b_lat} and \ref{fig:kappa_c} summarize the essential
features of the results for the heavy-quark transport coefficients. Notice in
particular how the dependence on the unphysical intermediate cutoff $|t|^*$ is
very mild. 
In these figures the weak coupling transport coefficients are compared 
to the ones from a lattice-QCD analysis, which provides results for $\kappa$ 
larger than the perturbative one by a sizeable factor a low momenta, where the
static approximation used in the lattice calculation is valid.
Lacking any information on their momentum dependence, we are forced to use the
same value also at large momenta: the growth with $p$ of the weak coupling
coefficients is such that in the transverse channel they become roughly of
the same size as the lattice-QCD one, whereas in the longitudinal channel they
become much larger. This fact has relevant consequences on the physical
observables, as we shall discuss in the following.

\subsection{Results}

\begin{table}[b]
\caption{Initial conditions for the hydrodynamical calculations simulating the
  background medium in $AA$ collisions at RHIC and the LHC. For the LHC, unless
  otherwise specified, we have employed, for the initial time, the standard
  value $\tau_0=0.6$~fm/c. As a test of the sensitivity on the thermalization
  time, in a few cases the extreme value $\tau_0=0.1$~fm/c has also been
  used.}
\label{tab:hydro}
\begin{center}
\begin{tabular}{|c|c|c|c|c|}
\hline
Nuclei & $\sqrt{s_{\rm NN}}$ & $\tau_0$ (fm/c) & $s_0$ (fm$^{-3}$) & $T_0$ (MeV)\\
\hline
Au-Au & 200 GeV & 1.0 & 84 & 333\\
\hline
Pb-Pb & 2.76 TeV & 0.6 & 278 & 475\\
\hline
Pb-Pb & 2.76 TeV & 0.1 & 1688 & 828\\
\hline
\end{tabular}
\end{center}
\end{table}

In this section we present the results of our Langevin setup, comparing them
with the recent experimental data obtained by the STAR, ALICE and CMS
collaborations at RHIC and the LHC. 
The medium produced in Au-Au and Pb-Pb collisions is described through
hydrodynamical calculations performed with the viscous 2+1 code of
Ref.~\cite{rom1}, using Glauber initial conditions with $\sigma_{NN}=42$~mb and 
$\sigma_{NN}=64$~mb for RHIC and the LHC, respectively.
The parameters used for the code initialization are summarized in
Table~\ref{tab:hydro}, whereas in Table~\ref{tab:b} we show the impact
parameters corresponding to various centrality classes (results in other
centrality classes have been obtained by a weighted sum of the cases of
Table~\ref{tab:b}). 

The initial hard $Q\overline{Q}$ production is simulated through the
POWHEG+PYTHIA setup described in Sec.~\ref{sec:pp}, supplemented in the $AA$
case by EPS09 \cite{Esk09} nuclear corrections to the PDFs. 
Transverse momentum broadening 
in nuclear matter is also introduced, according to the procedure described
in Ref.~\cite{lange}. We summarize the results for the total $c\overline{c}$ and
$b\overline{b}$ cross sections in Table~\ref{tab:cross}. 

\begin{table}[t]
\caption{\label{tab:b} The centrality classes and the corresponding average
  impact parameters at the kinematics of RHIC and the LHC.}
\begin{center}
\begin{tabular}{|rr|rr|}
\hline
    \multicolumn{2}{|c|}{Au-Au ($\sqrt{s}=200$ GeV)} & 
    \multicolumn{2}{c|}{Pb-Pb ($\sqrt{s}=2.76$ TeV)}    \\
\hline
   $C_1$-$C_2$  & $b$ (fm) & 
   $C_1$-$C_2$  & $b$ (fm) \\
\hline
  0-10\%  &  3.27 &  0-10\% &  3.32 \\
 10-20\%  &  5.78 & 10-20\% &  6.04 \\
 20-40\%  &  8.12 & 20-30\% &  7.82 \\
 40-60\%  & 10.51 & 30-40\% &  9.25 \\
 60-80\%  & 12.42 & 40-50\% & 10.63 \\
          &       & 50-60\% & 11.74 \\
          &       & 60-80\% & 13.12 \\
          &       & 80-100\% & 15.08 \\
\hline
\end{tabular}
\end{center}
\end{table}

\begin{table}[b]
\caption{Total $c\overline{c}$ and $b\overline{b}$ cross sections in $pp$ and
  $AA$ collisions at RHIC and the LHC, calculated with POWHEG using the default
  renormalization and factorization scales and the CTEQ6M PDF (supplemented by
  the EPS09 nuclear modifications in the case of $AA$ collisions); $m_c=1.3$~GeV
  and $m_b=4.8$~GeV.} 
\label{tab:cross} 
\begin{center}
\begin{tabular}{|c|c|c|c|}
\hline
Collision & $\sqrt{s_{\rm NN}}$ & $\sigma_{c\overline{c}}$ (mb) &
$\sigma_{b\overline{b}}$(mb) \\
\hline
$pp$ & 200 GeV & 0.405 & $1.77\times 10^{-3}$ \\
\hline
Au-Au & 200 GeV & 0.356 & $2.03\times 10^{-3}$ \\
\hline
$pp$ & 2.76 TeV & 2.425 & 0.091 \\
\hline
Pb-Pb & 2.76 TeV & 1.828 & 0.085 \\
\hline
\end{tabular}
\end{center}
\end{table}

The propagation in the plasma is then studied thro\-ugh the Langevin setup
described above. In the the HTL transport coefficients the strong coupling $g$
(for soft collisions), if not otherwise specified, is evaluated at the scale
$\mu=1.5\pi T$, representing the central value of the systematic band explored
in our study.

\begin{figure}
\begin{center}
\includegraphics[clip,width=0.48\textwidth]{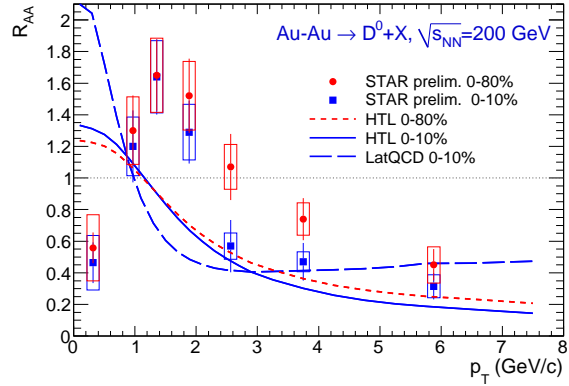}
\caption{Results (with HTL and lattice-QCD transport coefficients) for the
  $R_{AA}$ of $D^0$ mesons in central ($0-10\%$, in blue) and minimum-bias
  ($0-80\%$, in red) Au-Au collisions at $\sqrt{s_{NN}}=200$ GeV
  compared to preliminary STAR data~\cite{STAR_D}.}
\label{fig:RAA_STAR}
\end{center} 
\end{figure}

In Fig.~\ref{fig:RAA_STAR} we display the outcomes of our Langevin setup (with
weak coupling HTL transport coefficients as well as with lattice-QCD ones) for
the nuclear modification factors $R_{AA}(p_T)$ of $D^0$ mesons in central and
minimum-bias Au-Au collisions at RHIC, compared to preliminary STAR
data~\cite{STAR_D}. The size of the suppression for $p_T\gsim 2$ GeV in central
($0-10\%$) events is quite well reproduced by the HTL curve. On the other hand
experimental data display a bump around $p_T\sim 1.5$ GeV/c, with $R_{AA}>1$ in
the $p_T$ range 1--2 GeV/c and a depletion at smaller $p_T$, which is missed by
our model.

Such a non-trivial behaviour at low $p_T$ (say, for $p_T \lsim3$ GeV/c)
--- made visible by the very fine binning in the low-momentum region ---
might come from coalescence \cite{vanHees}, so far not implemented into our
framework: $D$ mesons in a given $p_T$-bin, in $pp$ collisions can only come
from the fragmentation of $c$ quarks with higher momentum; if on the other hand
coalescence were at work in the $AA$ case, the $D$ mesons in the same $p_T$ bin
might come from the much more abundant $c$ quarks at lower $p_T$, leading to
an enhancement of the spectrum. 
  
The $p_T$ behaviour of the quenching obtained with lattice-QCD transport
coefficients on the other hand looks quite different from the one of the weak
coupling calculations. In particular, the strong suppression of the
spectra at moderate $p_T$ leads to a steep rise of $R_{AA}$ for $p_T\!\to\!0$
not observed in the data: however, since, as we noticed, this is the region most
affected by a possible coalescence mechanism, one cannot draw any definite
conclusion. At larger momenta the lattice-QCD results show a mild growth of
$R_{AA}$, at variance with both the HTL calculations and the data, but this is
likely a consequence of the lack of any momentum dependence in the lattice-QCD
transport coefficients.

\begin{figure*}
\begin{center}
\includegraphics[clip,width=1.0\textwidth]{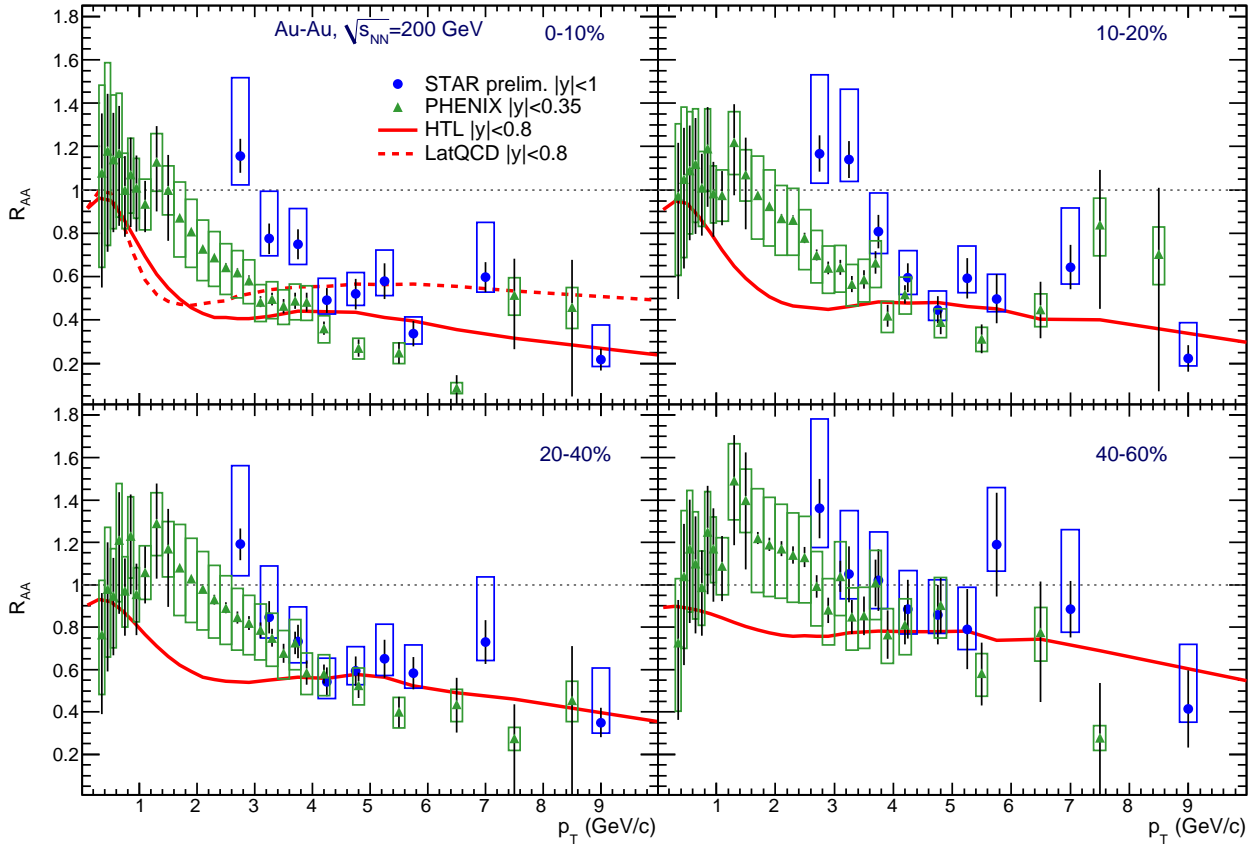}
\caption{Results (with HTL transport coefficients) for the $R_{AA}$ of
  non-photonic electrons ($e_{c+b}$) from charm and beauty decays in Au-Au
  collisions at $\sqrt{s_{NN}}\!=\!200$ GeV for various centrality classes
  compared to PHENIX~\cite{PHE} and preliminary STAR~\cite{STARhfe}
  data. In the upper left panel we include also the result with lattice-QCD 
  transport coefficients.}
\label{fig:RAA_e_STAR-PHENIX_4panel}
\end{center}
\end{figure*}

Notably --- at variance with the experimental data, which for moderate $p_T$
display a milder quenching in the $0-80\%$ centrality class --- Langevin results 
for minimum-bias collisions are very similar to the ones in more central events.
Actually, such a difference between central and minimum-bias events is not
apparent in the PHENIX data \cite{PHE} for non-photonic electrons (see
Ref.~\cite{lange} for a comparison with our model). Indeed, some amount of
discrepancy is present between these PHENIX data and the preliminary STAR data
for non-photonic electrons \cite{STARhfe}, as one can see in
Fig.~\ref{fig:RAA_e_STAR-PHENIX_4panel}, where we compare them to our Langevin
outcomes for heavy flavour decay electrons ($e_{c+b}$) in Au-Au collisions for
different centrality classes. 

\begin{figure*}
\begin{center}
\includegraphics[clip,width=0.48\textwidth]{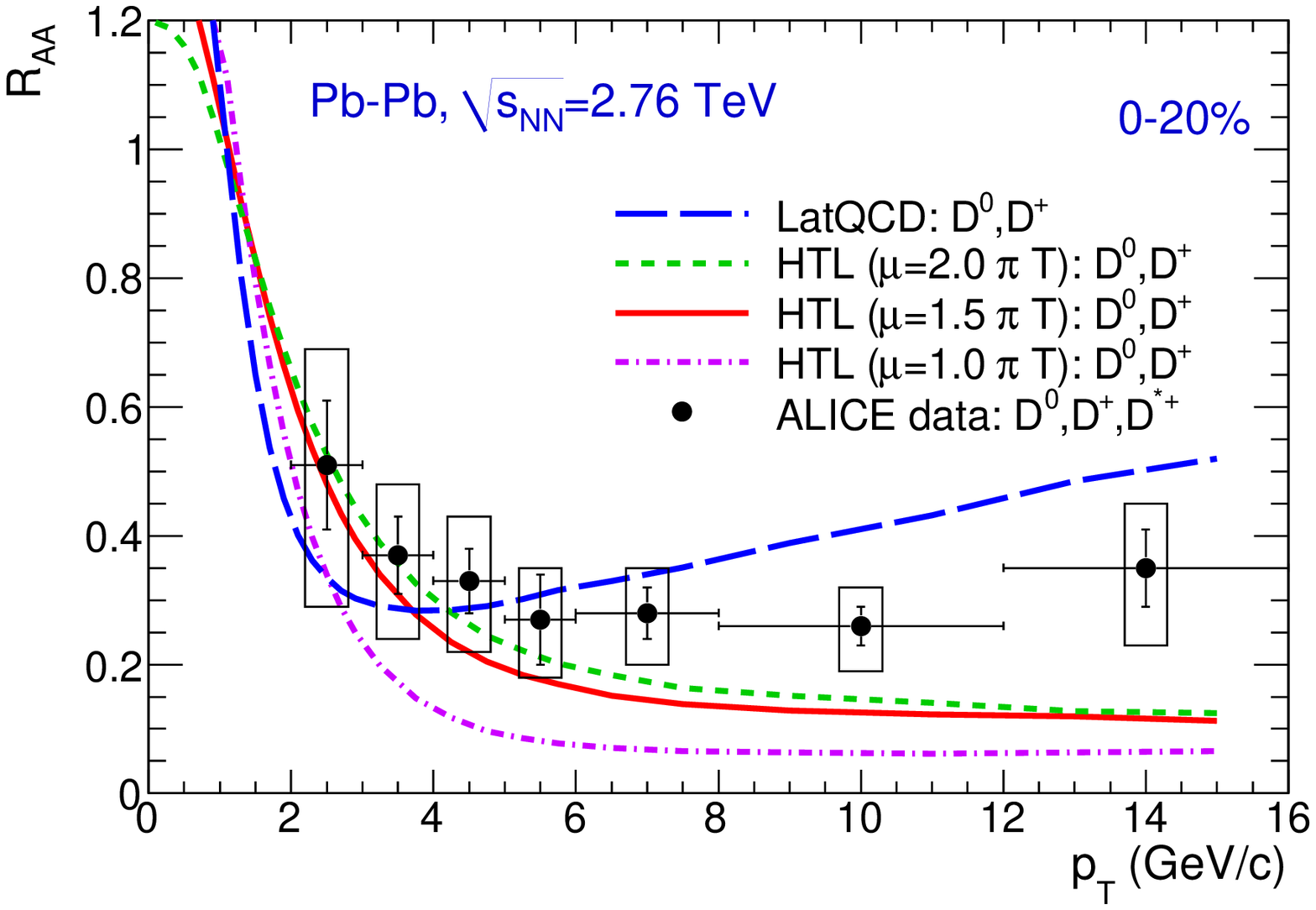}
\includegraphics[clip,width=0.48\textwidth]{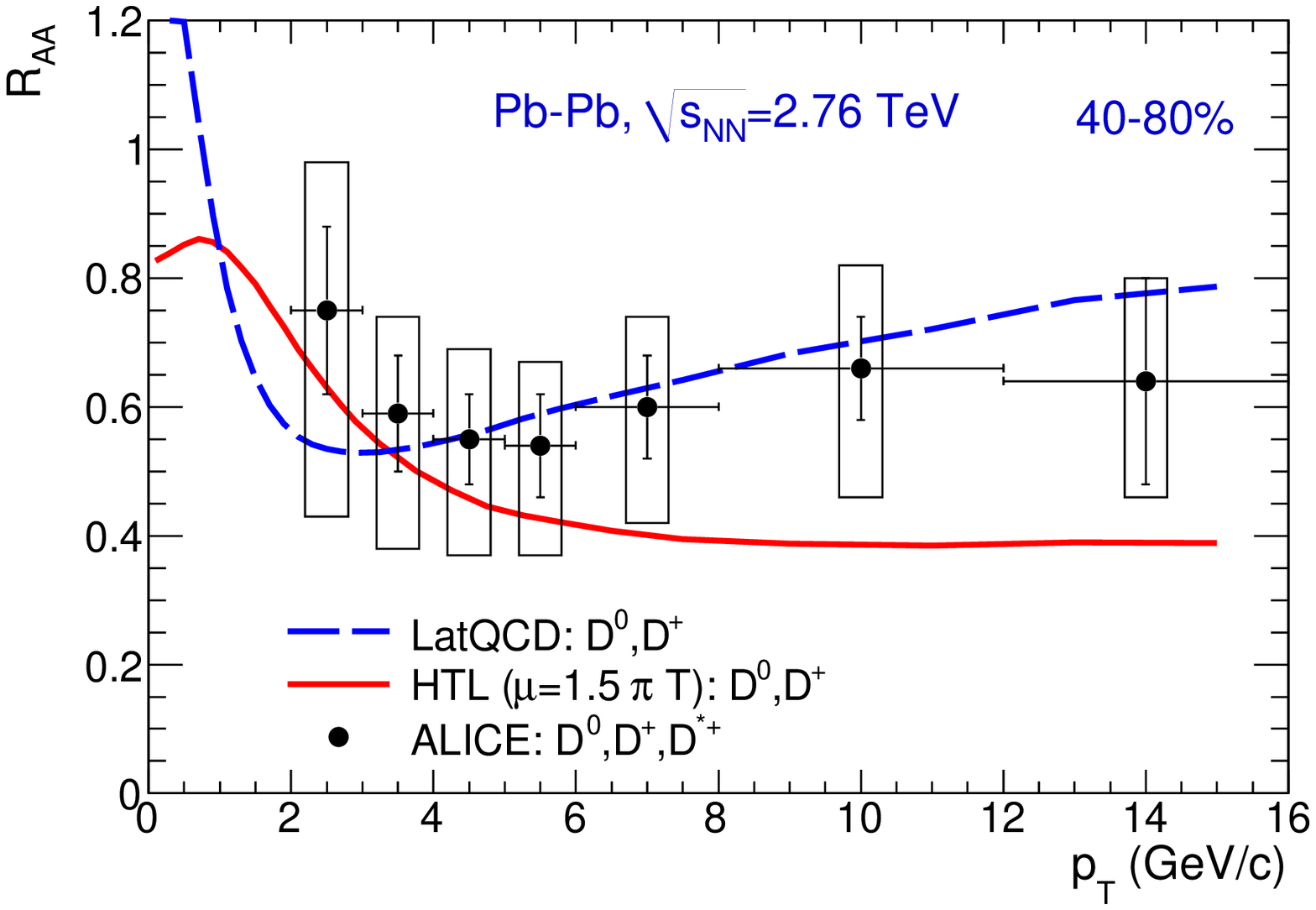}
\caption{A systematic study of the $D$-meson nuclear modification factor for
  different choices of the transport coefficients. In the HTL case we let the
  scale of the coupling vary from $\mu\!=\!\pi T$ to $\mu\!=\!2\pi T$. Results
  obtained with lattice-QCD transport coefficients are also shown for
  comparison. Our results are compared to ALICE data collected in the
  $0-20\%$ most central Pb-Pb collisions at $\sqrt{s_{NN}}\!=\!2.76$
  TeV~\cite{ALI_D} (left panel) and in semi-periphal events ($40-80\%$, right
  panel). }
\label{fig:RAA_D}
\end{center}
\end{figure*}

\begin{figure*}
\begin{center}
\includegraphics[clip,width=0.48\textwidth]{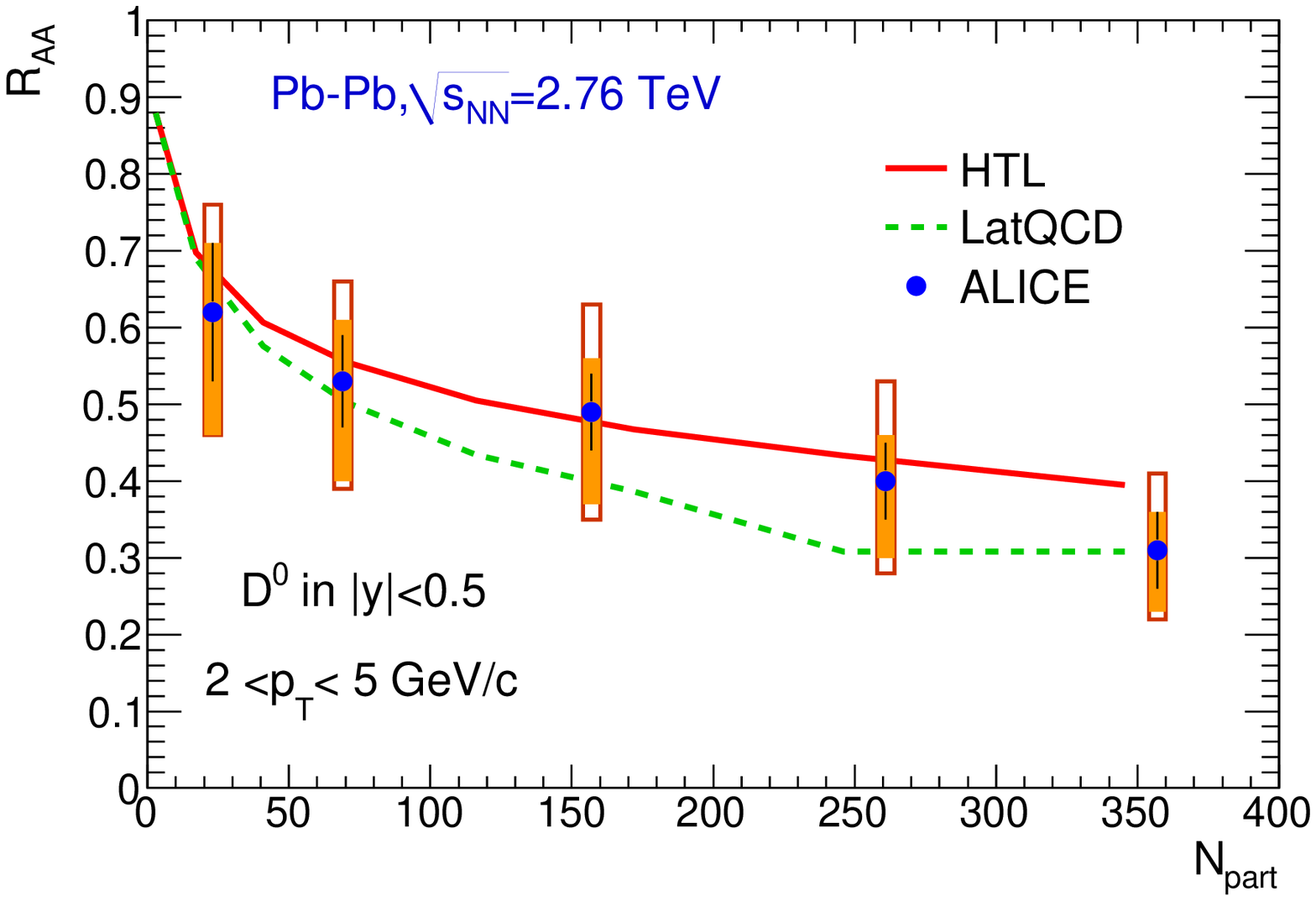}
\includegraphics[clip,width=0.48\textwidth]{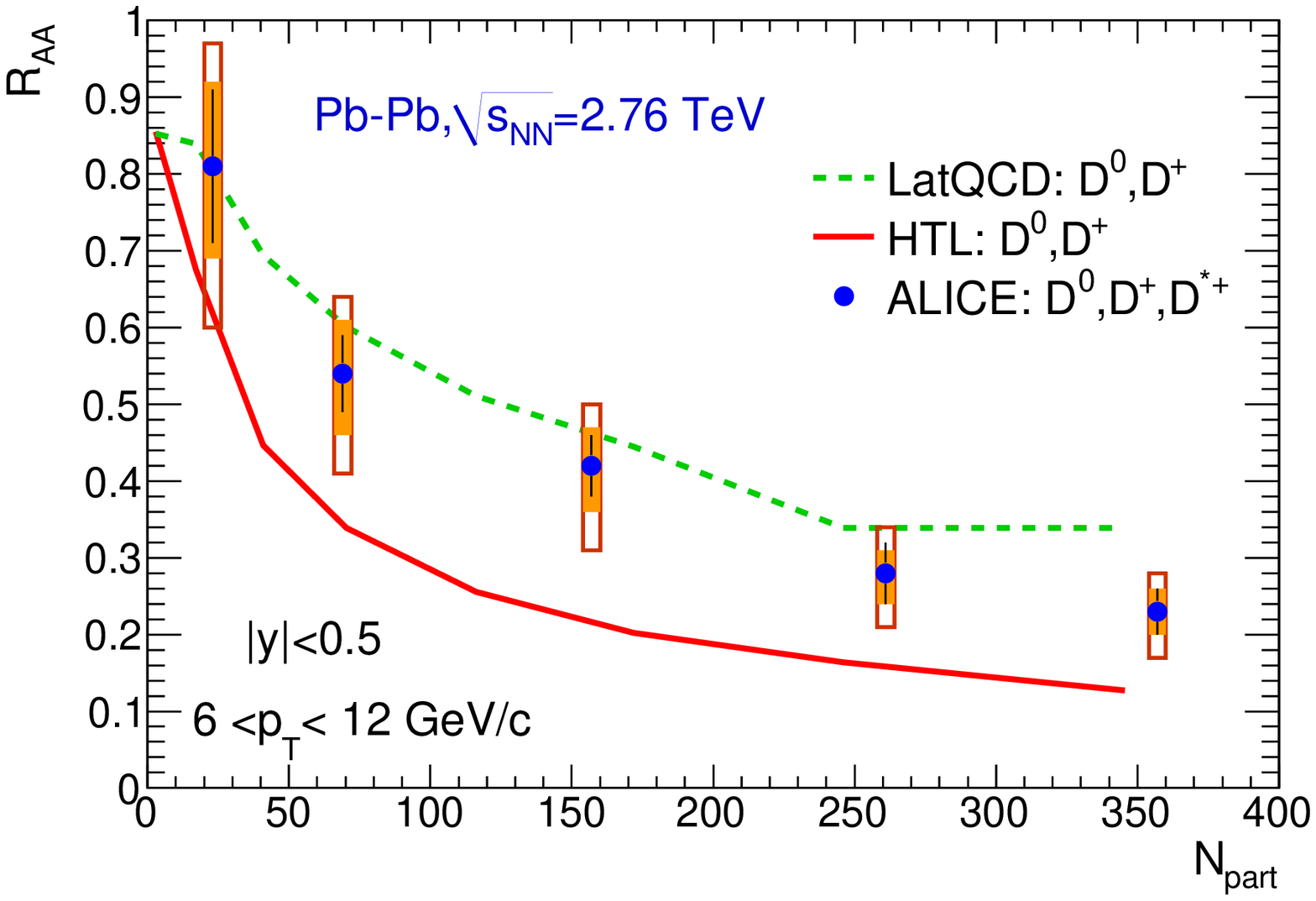}
\caption{Centrality dependence of the $D$-meson $R_{AA}$ in Pb-Pb
  collisions. Results with different transport coefficients are
  compared to ALICE data \cite{ALI_D} at moderate (left panel) and large (right
  panel) momenta.}
\label{fig:RAA_Dvscentr_highcut}
\end{center}
\end{figure*}

\begin{figure}[b]
\begin{center}
\includegraphics[clip,width=0.48\textwidth]{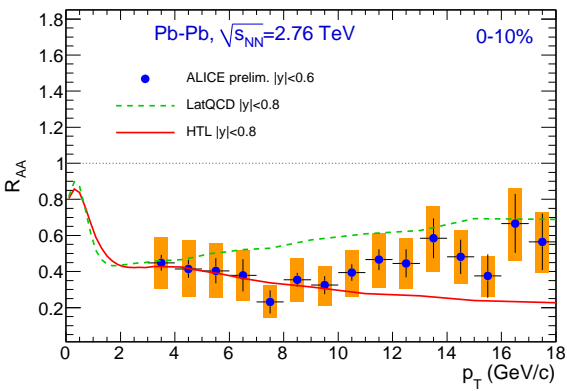}
\caption{Predictions (with HTL and lattice-QCD transport coefficients) for
  heavy-flavour decay electrons from charm and beauty ($e_{c+b}$) in Pb-Pb
  collisions at the LHC compared to preliminary ALICE experimental data
  \cite{Sakai} in $0-10\%$ most central events.}
\label{fig:RAA_e}
\end{center}
\end{figure}

Also in this case theory outcomes nicely reproduce the data for large
enough $p_T$ (say, $p_T\gsim 4$~GeV/c), missing the enhancement observed in the
low-momentum region. The present experimental systematic uncertainties and the
moderate discrepancies between the results by the two collaborations make
difficult to draw more definite conclusions.

In Figs.~\ref{fig:RAA_D}-\ref{fig:RAA_Dvscentr_highcut} we display our results
for the $D$-meson $R_{AA}$ (as a function of $p_T$ and of the centrality) in
Pb-Pb collisions at the LHC compared to ALICE data \cite{ALI_D}. In the LHC
case we performed a wider systematic study exploring the sensitivity of the HTL
results to the scale of the coupling. As shown in the left panel of
Fig.~\ref{fig:RAA_D}, HTL transport coefficients reproduce quite nicely the
data at moderate $p_T$, but at larger $p_T$ they would entail a too strong
quenching of the spectra, presumably due to the rapid rise of $\kappa_L(p)$
with the quark momentum. 
Note that the data seem to favour the larger values of the scale $\mu$ in the QCD
coupling and, interestingly, one observes a sort of saturation when increasing
the value of $\mu$, thus reducing the sensitivity of the results to this unknown
parameter. 

\begin{figure}
\begin{center}
\includegraphics[clip,width=0.48\textwidth]{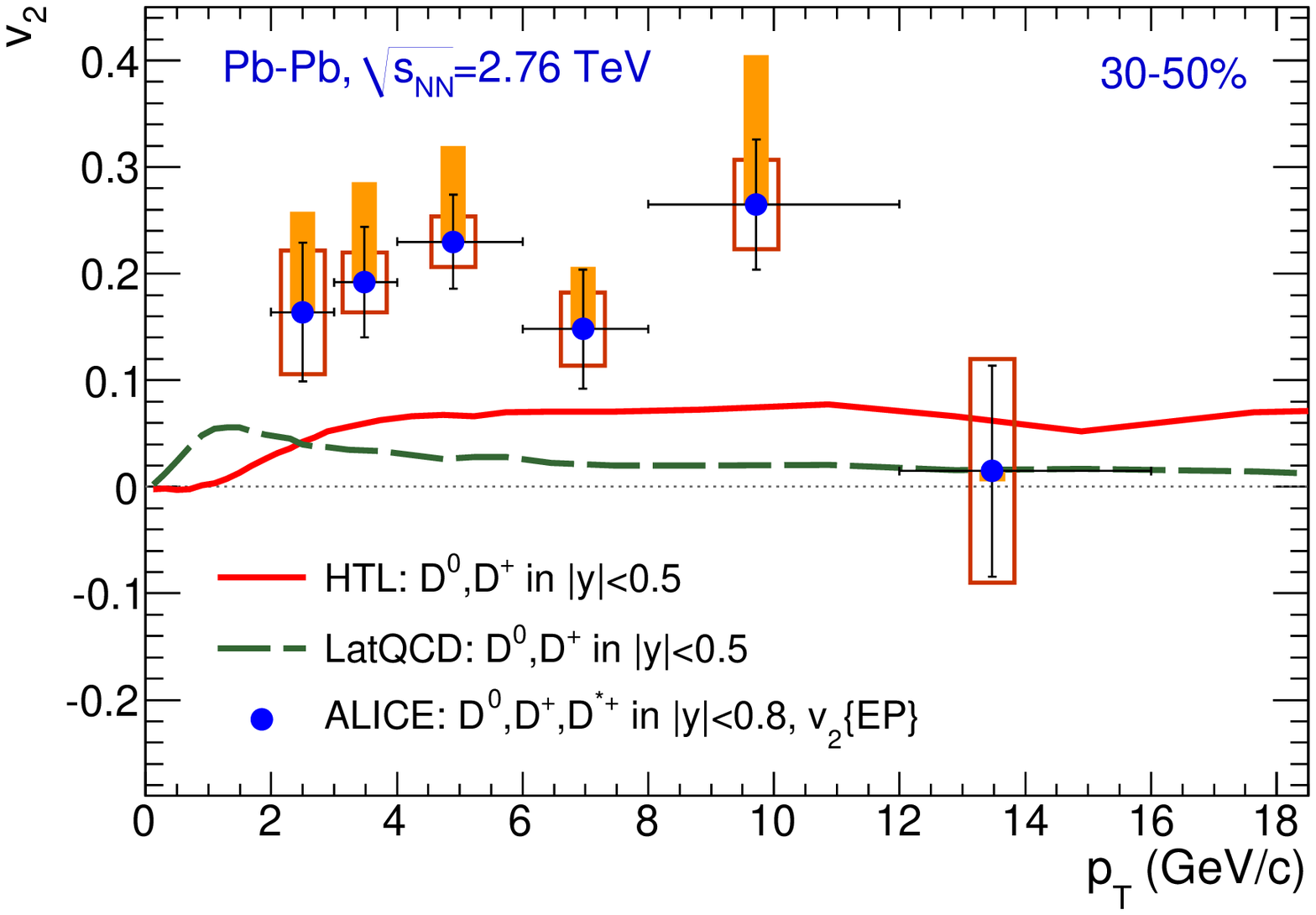}
\includegraphics[clip,width=0.48\textwidth]{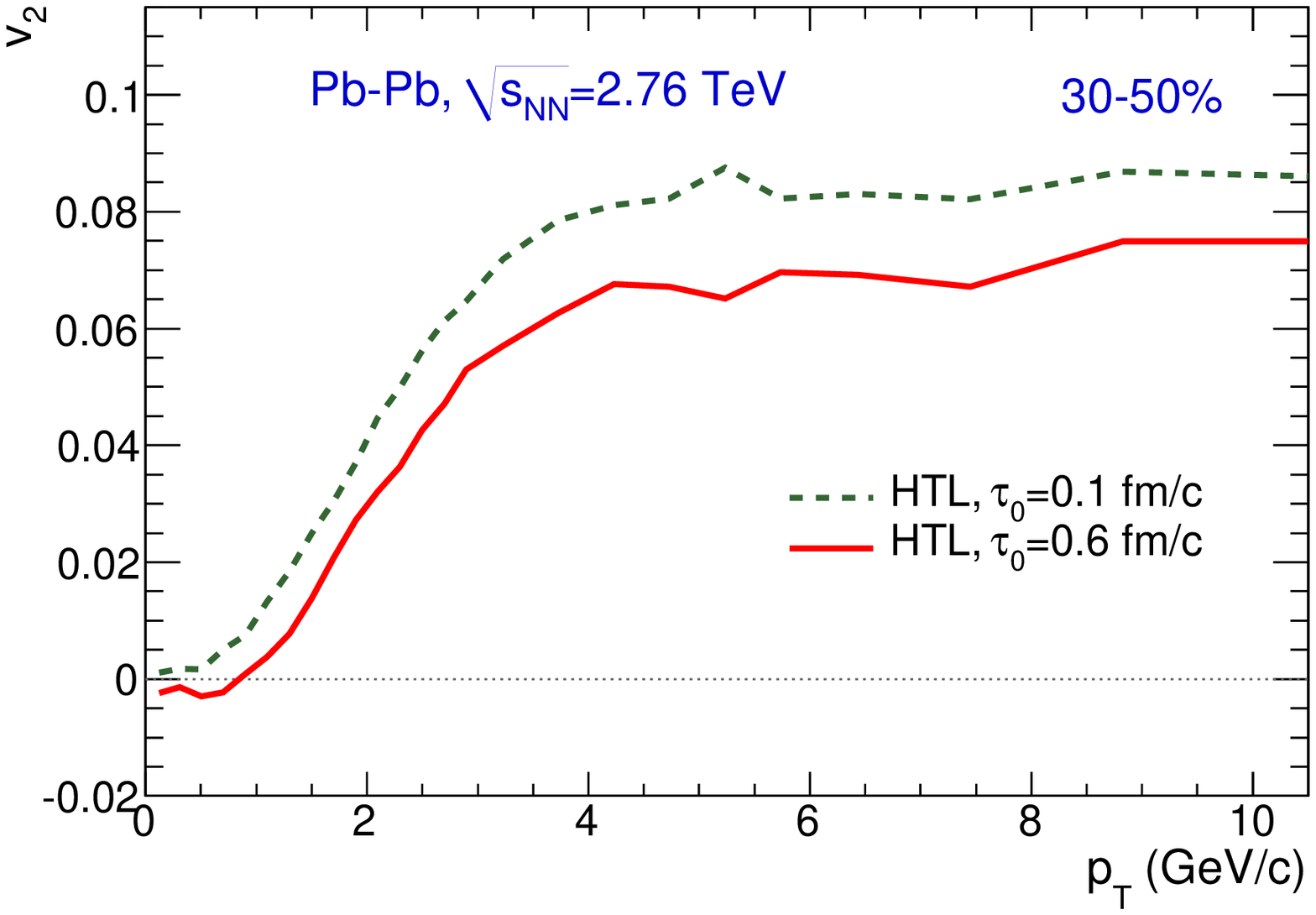}
\caption{Top panel: elliptic flow of $D$ meson in semi-peripheral
  ($30-50\%$ centrality class) Pb-Pb collisions at the LHC compared to
  ALICE data~\cite{ALIv2D}. Results obtained with HTL and
  lattice-QCD transport coefficients are displayed. 
  Bottom panel: dependence of the $D$-meson $v_2$ on
  the thermalization time $\tau_0$ of the background medium.}
\label{fig:v2_D}
\end{center}
\end{figure}

\begin{figure}
\begin{center}
\includegraphics[clip,width=0.48\textwidth]{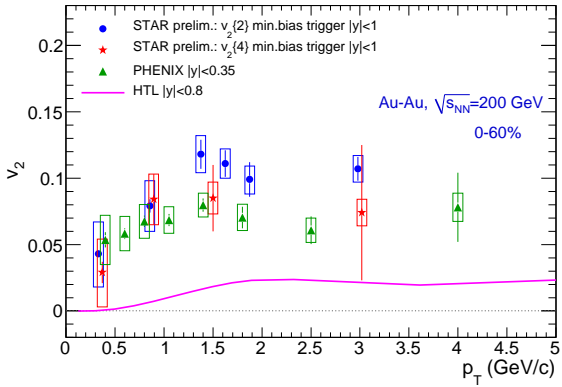}
\caption{Elliptic flow of non-photonic electrons: preliminary PHENIX
  \cite{PHE} and STAR \cite{STARhfe} data are compared to the outcomes of our
  Langevin calculations with HTL transport coefficients.}
\label{fig:v2e_STAR}
\end{center}
\end{figure}

On the other hand, ``lattice'' curves, obtained fixing
$\kappa$ (independent on the momentum) through the static results of
Ref.~\cite{hflat1}, display a too strong rise of $R_{AA}$ with $p_T$ compared
to data, suggesting that the actual behaviour stays probably in between, with a
mild dependence of $\kappa_L$ on the quark momentum.
This finding is also confirmed by the centrality dependence of $R_{AA}$ at high
momenta (right panel of Fig.~\ref{fig:RAA_Dvscentr_highcut}), whereas at
moderate momenta the two theoretical models give essentially the same
predictions (left panel of Fig.~\ref{fig:RAA_Dvscentr_highcut}).

Preliminary ALICE results \cite{Sakai} on electrons from charm and beauty
decays have also become available; we display them in Fig.~\ref{fig:RAA_e}
compared with the outcomes of our Langevin simulations for $R_{AA}$ in
central events. The size of the suppression is quite well reproduced.
  
Let us now consider the azimuthal anisotropy of the momenta of the heavy
flavour hadrons produced in the collision and of their decay electrons .
The anisotropy is characterized by the Fourier coefficients
$v_n=\langle\cos[n(\varphi-\Psi_{\rm RP})]\rangle$, where $\varphi$ is the
particle azimuthal angle and $\Psi_{\rm RP}$ is the azimuthal angle of the
reaction plane, which is defined by the impact parameter of the colliding
nuclei and the beam direction.
For non-central collisions, the dominant harmonic in the Fourier series is
the second one, $v_2$, commonly called elliptic flow, which reflects the
lenticular shape of the overlap region of the colliding nuclei.
Non-zero elliptic flow of final state hadrons originates from the build-up
of a collective motion of the medium constituents (dominant at low $p_{\rm
T}$)~\cite{ARTICOLODIFLOW} and from the path-length dependence of in-medium
parton energy loss.

\begin{figure}
\begin{center}
\includegraphics[clip,width=0.48\textwidth]{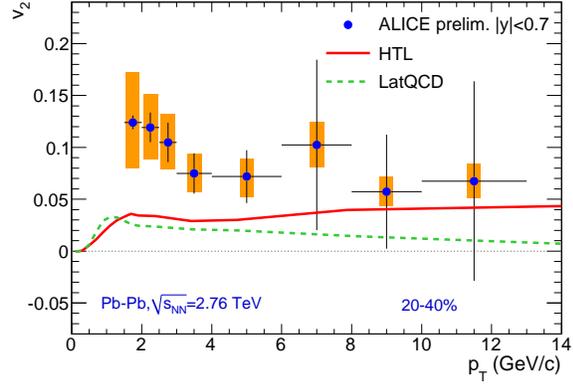}
\caption{Elliptic flow of non-photonic electrons: preliminary ALICE data
  \cite{Sakai} are compared to the outcomes of our Langevin calculations with HTL
  and lattice-QCD transport coefficients.}
\label{fig:v2e_ALICE}
\end{center}
\end{figure}

\begin{figure}[b]
\begin{center}
\includegraphics[clip,width=0.48\textwidth]{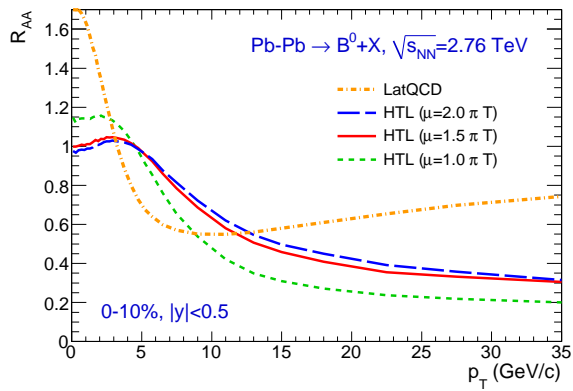}
\caption{Nuclear modification factor of $B^0$ mesons in central Pb-Pb
  collisions at the LHC provided by our setup for different
  choices of the transport coefficients (weak-coupling HTL calculations vs
  lattice-QCD simulations).}\label{fig:RAA_B}
\end{center}
\end{figure}

In Fig.~\ref{fig:v2_D} we address the elliptic flow of $D$ mesons. 
Outcomes of our Langevin setup for the elliptic flow $v_2$ are
compared to ALICE data \cite{ALIv2D} in semi-peripheral ($30\!-\!50\%$) Pb-Pb
collisions. HTL results significantly underestimate the experimental data at
low $p_T$, achieving at larger $p_T$'s an asymptotic plateau (experimentally
observed also in the case of light-hadron spectra) arising from the path-length 
dependence of the energy loss. Lattice results on the other
hand do a slightly better job at low $p_T$, but neglecting any possible energy
dependence of the momentum broadening coefficient $\kappa$ leads to underestimate
the flow at higher $p_T$. 

\begin{figure*}
\begin{center}
\includegraphics[clip,width=0.48\textwidth]{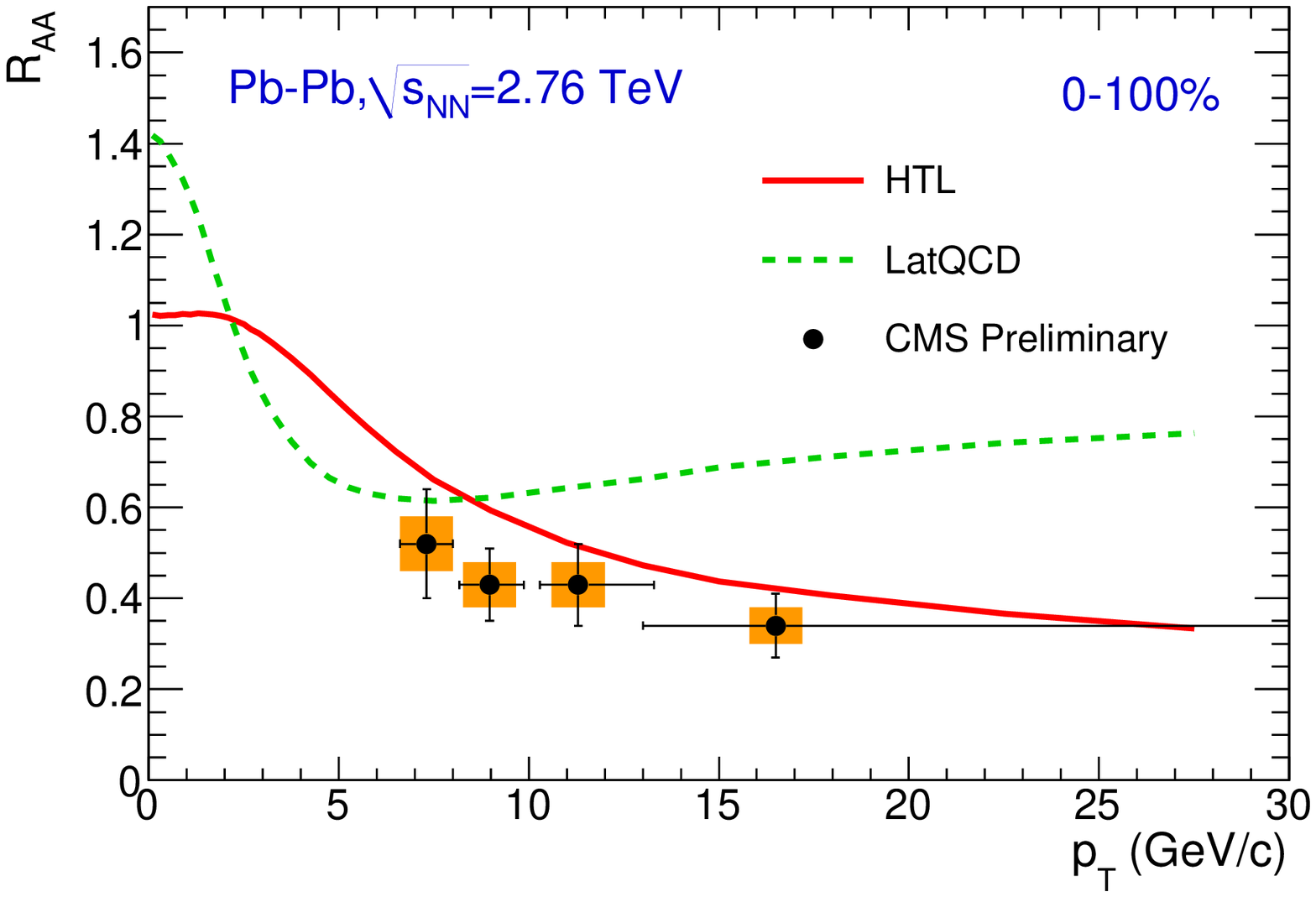}
\includegraphics[clip,width=0.48\textwidth]{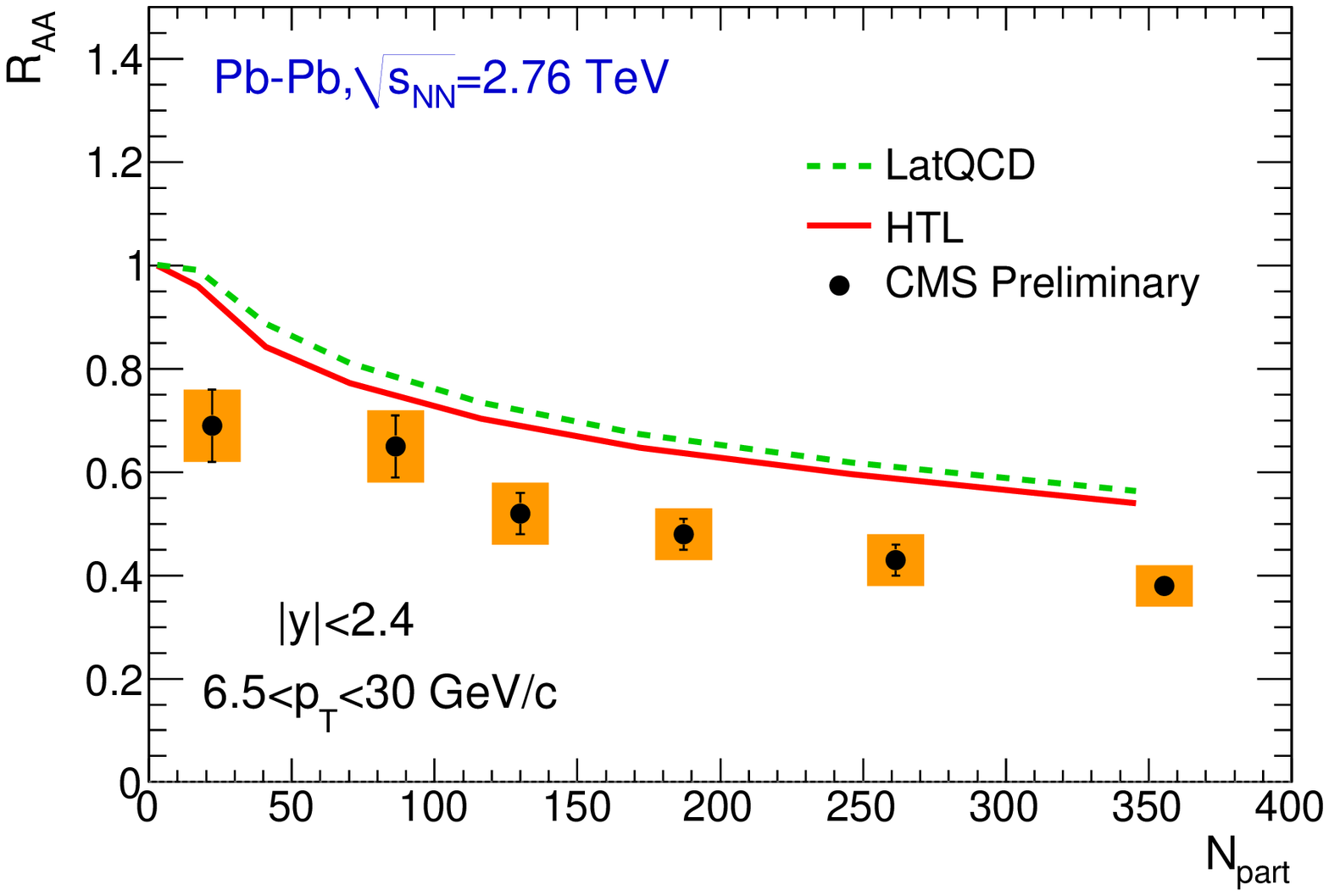}
\caption{Left panel: $R_{AA}$ as a function of $p_T$ of non-prompt $J/\psi$'s
  (from B decays) in minimum-bias Pb-Pb collisions at the LHC. 
  Results of our setup (with HTL and lattice-QCD transport coefficients) are
  compared to preliminary CMS data \cite{CMSJpsi}. Right panel: the centrality
  dependence of the $R_{AA}$ of non-prompt $J/\psi$'s.}
\label{fig:RAA_Jpsi}
\end{center}
\end{figure*}

The fact that, even with the quite large value of
$\kappa$ provided by lattice-QCD simulation, one still underestimates the charm
$v_2$ at small $p_T$ is a strong hint that an important contribution to the
elliptic flow of $D$ mesons may come from coalescence with thermal partons at
hadronization. Notably, even the quite extreme choice $\tau_0=0.1$ fm/c for the
initial thermalization time of the medium leads just to a moderate increase of
$v_2$ in the low-$p_T$ region, not sufficient to reproduce the amount of
anisotropy displayed by the experimental data.

In Figs.~\ref{fig:v2e_STAR} and~\ref{fig:v2e_ALICE} the elliptic flow of heavy
quarks is studied through the electrons from the semileptonic decays of charm
and beauty, $e_{c+b}$.
In this case the discrepancy between Langevin outcomes and experimental data at
small $p_T$ is even larger, whereas at the largest $p_T$'s accessible by
ALICE the agreement between the preliminary data \cite{Sakai} and the HTL
calculations is rather good.

Finally, we apply our setup to the study of beauty dynamics in the QGP. 
Measurements of $B$ mesons will become possible with the upgrade program
of the ALICE detector. We display our results for $B^0$ mesons in Pb-Pb
collisions in Fig.~\ref{fig:RAA_B}, where curves obtained with different sets of
transport coefficients (HTL and lattice-QCD) are shown.
Here, the low-$p_T$ region is especially interesting: for the reasons explained
above, in the case of bottom quarks one expects the coalescence mechanism to
alter the shape of $R_{AA}$ much less than in the case of charm quarks,
allowing one, hopefully, to clarify whether the strong transport coefficients
predicted by lattice-QCD are realistic.
 
Indirect information on beauty in heavy-ion collisions is however already
experimentally accessible, thro\-ugh the non-prompt $J/\psi$ (from
$B\to J/\psi + X$ decays) measurements by CMS \cite{CMSJpsi}. 
Our results for the displaced $J/\psi$ $R_{AA}$, versus $p_T$ and centrality,
are shown in Fig.~\ref{fig:RAA_Jpsi} and compared to the CMS preliminary
results.  
The data seem to point to a stronger quenching than predicted by
theory, at variance with the charm data, whose quenching is generally
overestimated at large $p_T$'s, at least in the HTL model.
Here, $R_{AA}$ as a function of $N_{\rm part}$ (shown in the right panel of
Fig.~\ref{fig:RAA_Jpsi}) does not show appreciable differences
between the HTL and the lattice-QCD models. On the other hand, the minimum-bias 
$R_{AA}$ as a function of $p_T$ (left panel of Fig.~\ref{fig:RAA_Jpsi}) shows a
fair agreement with the one of the HTL calculation, although, we stress it
again, no data are available at low $p_T$, where the lattice-QCD predictions
are more reliable.

\section{Conclusions and perspectives}\label{sec:concl}

In this paper we have shown a rich set of results provided by our transport
setup for the study of heavy quarks in the QGP. Theory outcomes have been
compared to the most recent experimental data collected at RHIC and the LHC
($D$ mesons, non-photonic electrons and displaced $J/\psi$'s) concerning the
quenching and the elliptic flow of charm and beauty in AA collisions. 
If the experimental heavy-flavour $R_{AA}$ can be reproduced reasonably well
over most of the $p_T$ range experimentally accessible, a consistent
description within the same setup of the elliptic flow of charm is still
lacking. 
In particular neither with perturbative nor with (the much larger) lattice-QCD 
transport coefficients we are able to reproduce the sizable elliptic flow of
$D$ mesons at low $p_T$. This suggests that in order to reproduce the
experimental data a modelling of the hadronization stage including the
possibility of coalescence --- in which the final $D$ mesons inherit part of
the anisotropy from the light thermal partons --- might be necessary. 

We have discussed in the text how $B$ mesons should be less affected by effects
arising from hadronization and how at the same time theory predictions (in
particular the ones provided by lattice-QCD) should be more reliable in the
case of beauty. $B$-meson measurements at low and moderate $p_T$, possible in
the near future thanks to the upgrade of the ALICE detector, have the
potentiality of providing information on the heavy-flavour transport
coefficients favored by the experimental data. 

So far, indirect information on the behaviour of beauty in the medium is
provided by preliminary CMS analysis of non prompt $J/\psi$'s from $B$ decays. 
Results for the corresponding $R_{AA}$ have been compared to the outcomes of
our setup with different sets of transport coefficients: a decent agreement has
been found in the HTL case; the available experimental data on the other hand
refer to a too hard $p_T$-range to make meaningful a comparison with the
lattice-QCD case (for which there is no information on the momentum dependence
of the transport coefficients).
 
A few important items remain to be addressed and are left for future work.
First of all, a modelling of coalescence, necessary in order to provide
predictions at low $p_T$. Secondly, extending the setup
to the forward-rapidity region, so that one can study also the rapidity
dependence of the various heavy-flavour observables and face also the
single-muon data measured by the ALICE experiment. This step would require to
interface our transport setup with the output of a full 3+1 hydrodynamic code,
which is currently under development \cite{Del13}.
Finally, addressing the study of more differential observables, like
$Q\overline{Q}$ correlations (recently attempted by other groups~\cite{marle}),
which are accessible to us thanks to the use of an event generator for the
initial heavy quark production. This kind of studies is starting to trigger the
interest of the experimental community.
 
\section*{Acknowledgments}
The authors would like to thank Mateusz Ploskon and Daniel Kikola for providing
them preliminary heavy-flavour data by ALICE and STAR and Torsten Dahms for
fruitful discussions. This research has been supported by the Italian Ministery
of University and Public Instruction, under National Project 2009WA4R8W.

\appendix

\section{Lattice transport coefficients}\label{app:kappa}

For the sake of self-consistency, in this appendix, following the 
steps of Ref.~\cite{sola}, we display how the heavy quark momentum-diffusion
coefficient $\kappa$ can be given a general quantum field theory definition and
how it can be expressed in terms of quantities accessible by lattice-QCD
simulations. 

One has to address the quite common situation of a ``system'' (the heavy quark)
coupled to an ``environment'' (the thermal bath of gluons and light quarks).   

We have shown in Eq.~(\ref{eq:force-force}) how $\kappa$ is related to the
following force-force correlator: 
\beq
D^>(t)\equiv\frac{1}{3}\langle F^i(t)F^i(0)\rangle_{\rm HQ}.\label{eq:D>}
\eeq
Analogously, one defines $D^<(t)\equiv({1}/{3})\langle F^i(0)F^i(t)\rangle_{\rm
  HQ}$. KMS boundary
conditions entail for their Fourier transforms
$D^<(\omega)=e^{-\beta\omega}D^>(\omega)$. One has then for the corresponding
spectral function: 
\beq
\sigma(\omega)\equiv D^>(\omega)-D^<(\omega)=(1-e^{-\beta\omega})D^>(\omega).
\eeq
The momentum-diffusion coefficient reflects the $\omega\!\to\!0$ limit of the
above spectral density. In fact: 
\beq
\kappa\equiv\int_{-\infty}^{+\infty}\!\!\!dt\, D^>(t)=D^>(\omega=0).
\eeq
Hence one has
\beq
\kappa=\lim_{\omega\to
  0}\frac{\sigma(\omega)}{1-e^{-\beta\omega}}=\lim_{\omega\to
  0}\frac{T}{\omega}\sigma(\omega),\label{eq:kappaspec}
\eeq
$\sigma(\omega)$ being the quantity extracted from lattice-QCD simulations. For
the latter one needs to consider the coupling of a (infinitely) heavy quark
with the colour field. The starting point is the $M\!\to\!\infty$ limit of the
NRQCD lagrangian 
\beq
{\cal L}=Q^\dagger(i\partial_0+gA_0)Q,
\eeq
with the non-relativistic fields obeying the anticommutation relation
\beq
\left\{Q_i(t,\x),Q_j^\dagger(t,\y)\right\}=\delta_{ij}\delta(\x-\y)\label{eq:anticomm}
\eeq
and the heavy-quark evolution being described the path-ordered exponential
$U(t,t_0)$:
\beq
Q_i(t)={\cal
  P}\exp\left[ig\!\int_{t_0}^t\!A_0(t')dt'\right]_{ij}\!\!Q_j(t_0)=U_{ij}(t,t_0)Q_j(t_0). 
\eeq
One needs then to define the expectation value in Eq.~(\ref{eq:D>})
\beq
\langle F^i(t)F^i(0) \rangle_{\rm HQ}\equiv\frac{\sum_s\langle s|e^{-\beta
    H}F^i(t)F^i(0)|s\rangle}{\sum_s\langle s|e^{-\beta
    H}|s\rangle}\label{eq:FFdef},
\eeq
which is taken over a thermal ensemble of states $|s\rangle$ of the environment
\emph{plus} one additional heavy quark, namely: 
\beq
\sum_s\langle s|\dots|s\rangle\equiv\sum_{s'}\int\!d\x\, \langle
s'|Q_i(-T,\x) \dots Q_i^\dagger(-T,\x)|s'\rangle. 
\eeq
Viewing the thermal weight $e^{-\beta H}$ as the imaginary-time translation
operator, so that 
\beq
Q(-T)e^{-\beta H}=e^{-\beta H}e^{\beta H}Q(-T)e^{-\beta H}=e^{-\beta
  H}Q(-T-i\beta), 
\eeq
and exploiting the anticommutation relation~(\ref{eq:anticomm}) one gets for
the HQ partition function appearing in the denominator of Eq.~(\ref{eq:FFdef})
{\setlength\arraycolsep{1pt}
\beqa
Z_{\rm HQ}&=&\sum_{s'}\int\!d\x\, \langle s'|Q_i(-T,\x) e^{-\beta
  H} Q_i^\dagger(-T,\x)|s'\rangle\nonumber\\ 
{}&=&V_{\rm PS}\sum_{s'}\langle s'|e^{-\beta
  H}U_{ii}(-T\!-\!i\beta,-T)|s'\rangle\nonumber\\ 
{}&=&V_{\rm PS}\langle{\rm Tr}\,U(-T\!-\!i\beta,-T)\rangle Z_0,
\eeqa}
where now the thermal average is taken over the states of the environment
only, with partition function 
$Z_0\equiv\sum_{s'}\langle s'|e^{\beta H}|s'\rangle$, and the phase space
volume arises from 
\begin{displaymath}
\int\!d\x\,\delta(\x-\x)=\int\!d\x\int\!d\p/(2\pi)^3=V_{\rm PS}.
\end{displaymath}
The numerator in Eq.~(\ref{eq:FFdef}) can be evaluated analogously starting
from 
\begin{multline}
\sum_s\langle s|e^{-\beta
  H}\F(t)\!\cdot\!\F(0)|s\rangle=\sum_{s'}\int\!d\x\!\int\!d\r\!\int\!d\r'\\ 
\times\langle s'|Q_i(-T,\x) e^{-\beta
  H}Q_j^\dagger(t,\r)g\E_{jk}(t,\r)Q_k(t,\r)\\ 
\times Q_l^\dagger(0,\r')g\E_{lm}(0,\r')Q_m(0,\r')Q_i^\dagger(-T,\x)|s'\rangle.
\end{multline}
One gets then
\begin{multline}
\langle F^i(t)F^i(0) \rangle_{\rm HQ}=\langle{\rm
  Tr}[U(-T\!-\!i\beta,t)gE^i(t)\\ 
\times U(t,0)gE^i(0)U(0,-T)]\rangle/\langle{\rm
  Tr}\,U(-T\!-\!i\beta,-T)\rangle\label{eq:FFreal}. 
\end{multline}
The above definition is the one used in Ref.~\cite{sola}, in which the AdS/CFT
correspondence allows the derivation of real-time quantities in strongly
coupled gauge theories (${\cal N}\!=\!4$ SYM).
However, in lattice-QCD one has to rely on lattice simulations carried on in
euclidean time. Eq.~(\ref{eq:FFreal}) has to be accordingly generalized. 
In Refs.~\cite{hflat1,hflat2} the authors evaluated then the following
euclidean electric-field correlator~\cite{lai}:
\beq
D_E(\tau)=-\frac{1}{3}\frac{\langle{\rm
    Tr}[U(\beta,\tau)gE^i(\tau)U(\tau)gE^i(0)]\rangle}{\langle{\rm
    Tr}U(\beta,0)\rangle} 
\eeq
and from the latter they extracted the spectral function $\sigma(\omega)$
entering in Eq.~(\ref{eq:kappaspec}) according to 
\beq
D_E(\tau)=\int_0^\infty\frac{d\omega}{2\pi}\sigma(\omega)\frac{\cosh[(\beta/2-\tau)\omega]}{\sinh(\beta\omega/2)}. 
\eeq

\end{document}